\documentclass[a4paper,fleqn,usenatbib]{mnras}

\usepackage{newtxtext,newtxmath}
\usepackage[T1]{fontenc}
\usepackage{ae,aecompl}
\usepackage{graphicx}	
\usepackage{amsmath}	
\usepackage{amssymb}	
\usepackage{enumitem}

\newcommand{\pa}{$\pi$\,Aqr}
\newcommand{\bz}{BZ\,Cru}
\newcommand{\kms}{km\,s$^{-1}$}

\title[\pa\ and \bz\ Variability]{Let there be more variability in two $\gamma$\,Cas stars\thanks{Based on spectra obtained with Aur\'elie at OHP and with FEROS and UVES at ESO, as well as on data collected with the {\it SMEI}, {\it BRITE}, and {\it TESS} space missions.}}

\author[Y. Naz\'e et al.]{Ya\"el~Naz\'e$^1$\thanks{F.R.S.-FNRS Senior Research Associate, email: ynaze@uliege.be}, Andrzej Pigulski$^{2}$, Gregor Rauw$^1$, and Myron Smith$^{3}$
\\
$^{1}$Groupe d'Astrophysique des Hautes Energies, STAR, Universit\'e de Li\`ege, Quartier Agora (B5c, Institut d'Astrophysique et de G\'eophysique), \\
All\'ee du 6 Ao\^ut 19c, B-4000 Sart Tilman, Li\`ege, Belgium\\
$^{2}$Instytut Astronomiczny, Uniwersytet Wroc\l{}awski, Kopernika 11, 51-622 Wroc\l{}aw, Poland\\
$^{3}$NSF OIR Lab, 950 N Cherry Ave, Tucson, AZ 85721, USA
}

\pubyear{2019}

\begin{document}
\label{firstpage}
\pagerange{\pageref{firstpage}--\pageref{lastpage}}
\maketitle

\begin{abstract}
  We investigate the short-term optical variability of two $\gamma$\,Cas analogs, \pa\ and \bz, thanks to intensive ground-based spectroscopic and space-borne photometric monitorings. For both stars, low-amplitude (mmag) coherent photometric variability is detected. The associated signals display long-term amplitude variations, as in other Be stars. However, these signals appear at high frequencies, especially in \pa, indicating p-modes with a high degree $l$, a quite unusual feature amongst Be stars. While \bz\ presents only low-level spectral variability, without clear periodicity, this is not the case of \pa. In this star, the dominant photometric frequencies, near $\sim$12\,d$^{-1}$, are confirmed spectroscopically in separate monitorings taken during very different disk activity levels ; the spectroscopic analysis suggests a probable tesseral nature for the mode. 
\end{abstract}


\section{Introduction}
$\gamma$\,Cas, the first Be star identified, was long considered as a prototype of this category. However, the advent of high-energy facilities unveiled the presence of an intriguing peculiarity: very hard and moderately strong thermal X-ray emission \citep[for a review see \citealt{smi16}]{jer76,mas76}. In recent years, it has been realized that $\gamma$\,Cas is not an isolated case, with more than twenty ``$\gamma$\,Cas analogs'' currently known \citep{smi16,naz18}. All share the same high-energy properties: high plasma temperature ($kT\sim5-20$\,keV), X-ray luminosities intermediate between those of ``normal'' massive stars and those of X-ray binaries ($\log[L_{\rm X}/L_{\rm BOL}]$ between --6 and --4 and $L_{\rm X}[{\rm in\,2-10\,keV}]>10^{31}$\,erg\,cm$^{-2}$\,s$^{-1}$), as well as the presence of both short- and long-term variations in that energy range.

The origin of this peculiar X-ray emission is still under debate, with two main scenarios under consideration: accretion onto a compact object (a white dwarf, see e.g. \citealt{ham16,tsu18}, or a neutron star in a propeller regime, see e.g. \citealt{pos17}), or star-disk interactions \citep[and references therein]{smi16}. To help solve this 40\,yr-long mystery, all aspects of $\gamma$\,Cas stars need to be examined in detail. In this context, we decided to focus on a poorly-known property of those stars: their rapid variability in the optical domain. Such changes may directly inform the physical or dynamic state of the stellar photosphere. They can thus help us assess whether $\gamma$\,Cas analogs display distinct properties from other Be stars, in addition to their differences at high-energies.

In this paper, we examine ground-based spectroscopy and space-based photometry of the two visually brightest analogs after $\gamma$\,Cas itself: \pa\ and \bz\ (HD\,110432). After introducing the observations and their reduction in Section 2, Section 3 presents the results obtained for \pa\ and \bz, from which a general discussion follows in Section 4. A summary concludes the paper in Section 5.

\section{Observations and data reduction} 

\subsection{Optical spectroscopy}

\begin{table*}
\centering
\caption{Characteristics of the spectroscopic datasets.}
\label{opt}
\setlength{\tabcolsep}{3.3pt}
\begin{tabular}{lcccccccl}
  \hline\hline
Tel.+Inst. & Obs. ID & Date & \# of spec & range (\AA) & Res. power & $SNR$ & Exp. Time (s) & Detector \\
\hline
\multicolumn{9}{l}{\pa}\\
2.2\,m MPG/ESO+FEROS & 077.D-0390 & 2006 June 5--6 & 41 & 3600--7000 & 48000 & 465 & 450 & {\scriptsize EEV CCD (2048$\times$4096 of 15\,$\mu$m px)} \\
1.52\,m OHP+Aur\'elie&            & 2019 August 22--24 & 42 & 4440--4890 & 7000 & 535 & 600 & {\scriptsize Andor Newton 940 CCD (512$\times$2048 of 13.5\,$\mu$m px)} \\
\vspace*{-0.2cm}\\
\multicolumn{9}{l}{\bz}\\
ESO UT2+UVES & 071.C-0367 & 2003 June 15--16 & 53 & 3800--5000 & 100000 & 450 & 80 & $^a$  \\
             & 082.C-0566 & 2009 March 8 & 10 & 3050--3850 & 100000 & 240 & 60 & $^a$ \\
             &            &              &+10 & 4600--6650 &        & 450 & 60 &  \\
             &            &              & 10 & 3800--5000 &        & 325 & 60 &  \\
             &            &              &+10 & 6700--10400&        & 210 & 60 &  \\
             & 194.C-0833 & 2018 June 18 & 10 & 3050--3850 & 100000 & 265 & 120 & $^a$ \\
             &            &              &+10 & 4600--6650 &        & 430 & 120 & \\
             &            &              &  8 & 3800--5000 &        & 410 & 120 & \\
             &            &              &+12 & 6700--10400&        & 205 & 120 & \\
\hline
\multicolumn{9}{l}{\scriptsize $^a$ blue arm: EEV CCD ; red arm: one CCD as in blue arm plus one MIT CCD (all 2048$\times$4096 of 15\,$\mu$m px)}\\
\end{tabular}
\end{table*}

Table \ref{opt} provides an overview of the spectroscopic observations.

For the 2006 FEROS data of \pa, we downloaded the 42 raw data from the ESO raw data archives\footnote{http://archive.eso.org/eso/eso\_archive\_main.html} and reduced them in a standard way using the FEROS context under MIDAS (version 17FEBpl 1.2), which also automatically accounts for the barycentric correction. Absorption by telluric lines near H$\alpha$ was corrected within IRAF using the template of \citet{hin00}. All high-resolution spectra were finally normalized over the same set of continuum windows using polynomials of low order. While most (36) spectra have $SNR>400$, one spectrum had a poor quality (signal-to-noise ratio of only 10) hence it was discarded. The OHP spectra of \pa\ were reduced in a standard way using MIDAS and normalized in the same way as the FEROS data. The barycentric correction was applied afterwards.

In parallel, \bz\ was monitored four times with UVES. The 2010 dataset (program 085.C-0799) covered only the near-UV, 3050--3850\,\AA, range, which is not well adapted for our needs hence these data were discarded. The 2003 dataset was downloaded from the ESO raw data archives and reduced in a standard way using ESO reflex. Since these are slicer data, the end products were 2D spectra whose rows were averaged to get the final stellar spectra. Unfortunately, many spectra were affected by saturation: only 53 spectra could be used to study weak emissions or absorption lines such as He\,{\sc i}\,$\lambda$4471\,\AA; the strong H$\beta$ emission was usable in only 4 spectra, prohibiting its use for a high-cadence variability study as ours. The 2009 and 2018 reduced spectra were directly downloaded from the science portal of ESO archives\footnote{http://archive.eso.org/scienceportal/}. They were taken in two different dichroic modes. The only important stellar lines in common between these two settings (thereby allowing for a longer time base) are H$\beta$ and He\,{\sc i}\,$\lambda$4921\,\AA, though the latter line suffers from imperfect order merging on its red side. The strong H$\alpha$ emission suffers from saturation effects hence was not usable. All high-resolution spectra were finally normalized over the same set of continuum windows using polynomials of low order. The barycentric correction, not included in the automatic processing, was applied afterwards.

\subsection{Optical photometry}

Broad-band optical photometry of \pa\ and \bz\ was collected for eight years with the Solar Mass Ejection Imager ({\it SMEI}, \citealt{eyl03,jac04}). This instrument was installed onboard the {\it Coriolis} spacecraft which was launched on 2003 January 6 on a Sun-synchronous orbit. It was composed of three white light cameras, each covering a $60^{\circ}\times3^{\circ}$ field-of-view. It continuously took 4\,s exposures and covered the entire sky in a single orbit of 102\,min duration. Without any filter, its sensitivity was primarily dictated by its CCD detector: the quantum efficiency peaked at 7000\AA\ but was above 10\% in the 4500--9500\AA\ domain \citep{eyl03}.

\begin{figure*}
  \begin{center}
\includegraphics[width=8cm,bb=3 3 725 725, clip]{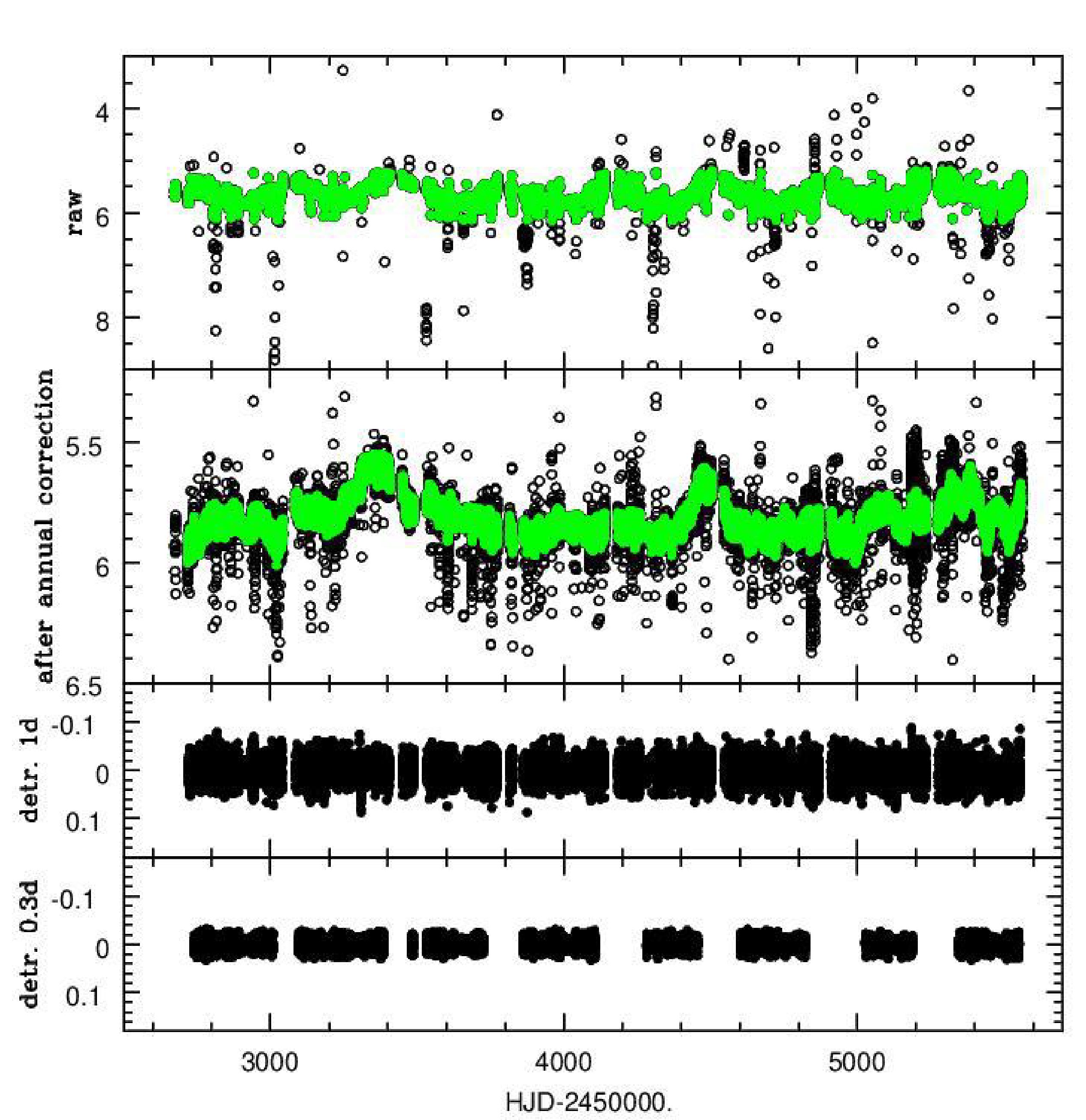}
\includegraphics[width=8cm,bb=3 3 725 725, clip]{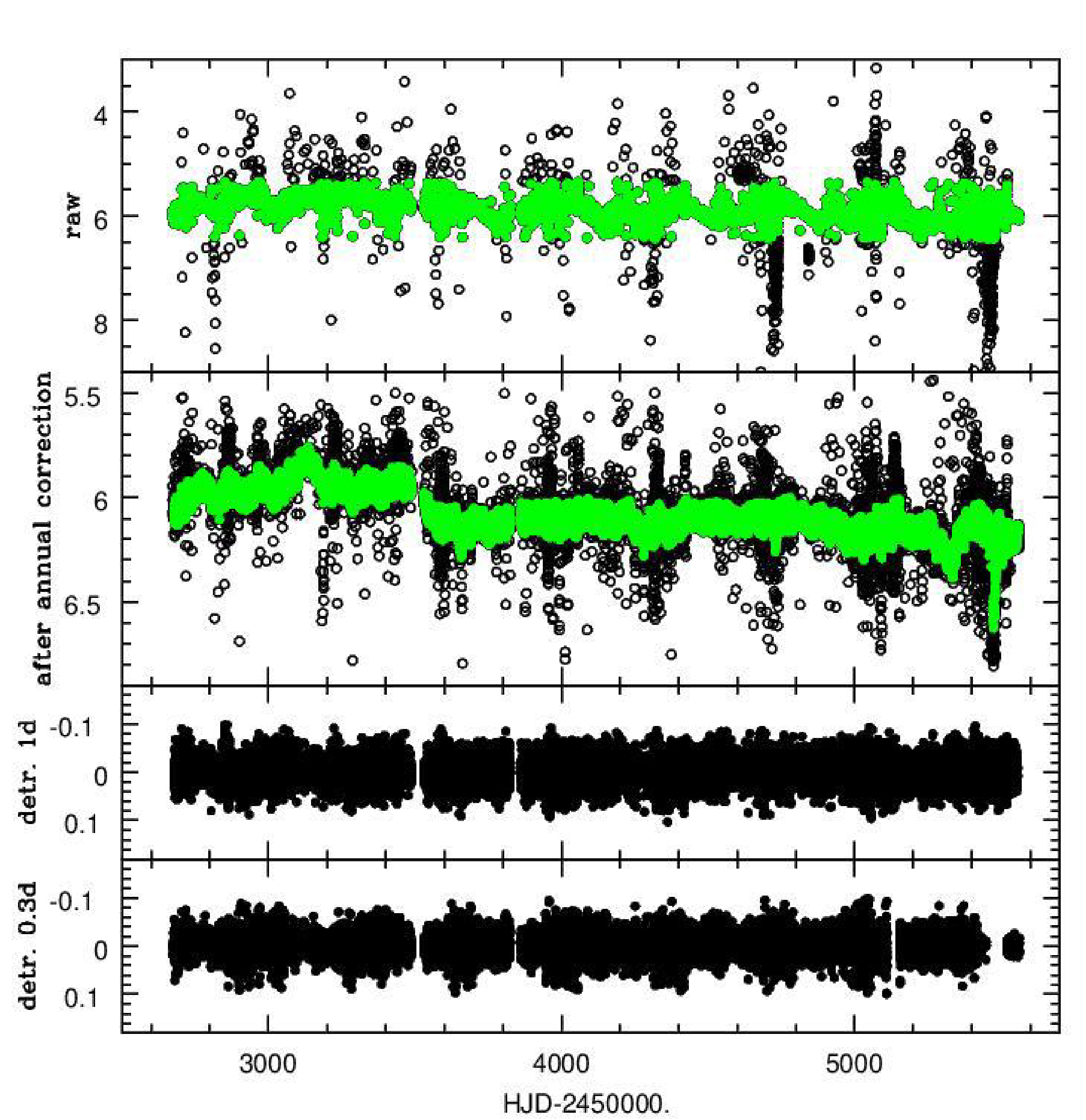}
  \end{center}
\caption{{\it SMEI} light curves at various stages of processing, for \pa\ on the left and \bz\ on the right. Ordinates are {\it SMEI} magnitudes (top two panels) or residual {\it SMEI} magnitudes (bottom two panels). {\it Top panels:} Raw light curves (black circles), with data points passing the 5MAD filtering in green. The yearly modulation can be seen.  {\it Middle upper panels:} Light curves after removing the yearly modulation (black circles), with data points passing the 3MAD filtering in green. {\it Bottom two panels:} Light curves after 1\,d or 0.3\,d detrending. }
\label{smei}
\end{figure*}

The primary goal of {\it SMEI} was to study the electron densities of the inner heliosphere on large scales by recording Thomson-scattered sunlight. To this aim, the contribution of zodiacal light and discrete background sources (stars brighter than 6th magnitude and solar system objects) had to be determined and subtracted. {\it SMEI} therefore provided stellar light curves as a byproduct. 

All {\it SMEI} images were processed in the same way \citep{hic07,hou10}: subtraction of dark current and an electronic (pedestal) offset, flat-field correction, identification and removal of ``hot'' pixels, energetic particles (trapped in the magnetosphere or from cosmic rays), and space debris. The data from each orbit were then combined and projected onto a sky grid with 0.1$^{\circ}$ spatial resolution. Finally, the undesired signals were taken out, using PSF-fitting for point sources. This led to the production of stellar light curves with a temporal resolution of 102\,min for each camera. These pipeline-processed light curves are available from the {\it SMEI} website\footnote{http://smei.ucsd.edu/new\_smei/data\&images/stars/timeseries.html}. 

The {\it SMEI} Point Spread Function (PSF) was strongly asymmetric, with extended shallow wings and a full width of about 1$^{\circ}$. Fortunately, \pa\ and \bz\ have no close ($<40^\prime$) neighbour of similar brightness: the brightest neighbour of \pa\ is $\Delta(V)$=3.9\,mag fainter and is located at 39$^\prime$, that of \bz\ has $\Delta(V)$=2.8\,mag and is located at 36$^\prime$. Furthermore, within 40$^\prime$ of \pa, the variable stars from the Variable Star index (VSX)\footnote{https://www.aavso.org/vsx/} catalog are fainter than the 15th\,mag while, for \bz, all VSX objects with $\Delta(V)$=2--3\,mag appear beyond a 30$^\prime$ separation. Therefore, we expect little contamination of the stellar signals.

{\it SMEI} nearly continuously observed \pa\ and \bz\ from January 2003 to December 2010, with only small 1-month gaps due to solar conjunctions. The raw {\it SMEI} light curve however presents numerous outliers (occurring singly and randomly) and instrumental variations, notably a strong yearly modulation \citep[e.g.][]{baa18}. To correct for those, we independently applied two cleaning techniques, allowing for cross-checks. Figure \ref{smei} shows the {\it SMEI} light curves at various stages of these processings.

The first cleaning method begins by rejecting points deviating from the median by more than 5 times the median absolute deviation (MAD). Then the data were phased with the sidereal year (365.25636\,d) and the yearly modulation was evaluated by taking the median of data in 100 phase bins. Outliers however remained even after correcting for this effect. Therefore a second rejection was needed. To this aim, a moving box smoothing with a 10\,d window allowed to determine the long-term trend of the data and points deviating from this trend by more than 3 MAD were removed. In the end, about 83\% of the initial data points were kept. Finally, the long-term trend was recalculated using a moving box smoothing with a 1\,d window and this trend was taken out. 

The second method performed a correction for the yearly modulation in a similar way as above, then the outliers were removed using a Generalized Extreme Studentized Deviate (GESD) algorithm \citep{ros83} with parameter $\alpha=0.3$ and samples of 250 consecutive data points. For such cleaned data, uncertainties of each data point were calculated from the scatter of nearby points. The uncertainties were calculated using residuals from the fit of a model consisting of nine sinusoidal terms with low frequencies. The model was not intended to have a physical meaning -- it was only used to account for the strongest intrinsic variability. Then, the worst data were removed by rejecting points with errors exceeding a user-defined threshold. The threshold was set lower and lower during subsequent iterations; the final one was equal to 0.02\,mag for \pa\ and 0.04\,mag for \bz. The procedure of the removal of the outliers and of the worst data was done iteratively. After each iteration the errors were recalculated and the 9-term model was fit to the data that remained. In the end, 52 and 76\% of the initial data points were kept for \pa\ and \bz, respectively. Finally, to reveal high-frequency signals, these cleaned data were detrended by calculating averages in 0.3\,d intervals, interpolating between them with Akima interpolation and subtracting the interpolated smooth curve. 

In addition to the {\it SMEI} data, \bz\ was also observed by one of the five nanosatellites composing the BRIght Target Explorer ({\it BRITE}, \citealt{wei14,pab16}) constellation: the {\it BRITE}-Toronto satellite, launched on 2014 June 19. This satellite uses a 3\,cm refractor equipped with a red filter, yielding a 5500--6900\,\AA\ passband. The field-of-view has a $\sim12^{\circ}$ radius and is observed every 20\,s during at least 15\,min per orbit (which has 98\,min duration). Considering the defocussing, the PSF has a 5\arcmin\ width, hence no contamination is expected for \bz. The observations of \bz\ took place in 2016 Feb--May during a run dedicated to the Crux-Carina\,I field; the data were downloaded from the public archives\footnote{http://brite.camk.edu.pl/pub/index.html}. The 4\,s exposures were taken in the chopping mode \citep{pab16}. Aperture photometry of the star was obtained from the pipeline described by \cite{pop17} and corrected for instrumental effects according to the procedure presented by \cite{pig18}. This procedure includes iterative one- and two-dimensional decorrelations, rejection of outliers and of the worst orbits. 

Finally, \bz\ was also observed by the Transiting Exoplanet Survey Satellite ({\it TESS}, \citealt{ric14,ric15}) in 2019 April--May (sector 11). Launched on 2018 April 18, {\it TESS} is placed on an ellipical, lunar synchronous, 13.7\,d orbit around the Earth. Equipped with a 10\,cm lens, {\it TESS} has a $>$90\% efficiency bandpass of 6000--9000\,\AA, a range including the Cousins $I$ band but broader than it. It also has a 50\% ensquared-energy half-width of 21\arcsec\ (corresponding to one detector pixel): no contamination from nearby stars is therefore expected for \bz. Its detector is composed of four mosaics with 2$\times$2 MIT CCDs which together cover a $24^{\circ}\times96^{\circ}$ strip. Images covering this wide field are taken every 30\,min, but subarrays on preselected stars (and \bz\ is one of them) are read every 2\,min. The data reduction pipeline is based on the {\it Kepler} mission pipeline, and it performs pixel-level calibration, background subtraction, flat-fielding, and bias subtraction. The {\it TESS} light curve of \bz\ was downloaded from the Mikulski Archive for Space Telescopes\footnote{https://mast.stsci.edu/}. Both simple aperture photometry as well as time series corrected for crowding, limited size of the aperture, and instrumental systematics are available. As the former light curve exhibits long-term trends, we decided to use only the latter light curve in this paper. Data points with a non-zero {\it TESS} quality flag were further discarded. The {\it TESS} fluxes were finally transformed into magnitudes using $mag=-2.5\times\log(flux)+21.5$ (the constant being arbitrarily chosen to have the resulting magnitudes near six).

\section{Results}
\subsection{\pa}

\subsubsection{Photometric analysis}

\begin{figure}
  \begin{center}
\includegraphics[width=8.5cm,bb= 3 3 715 750,clip]{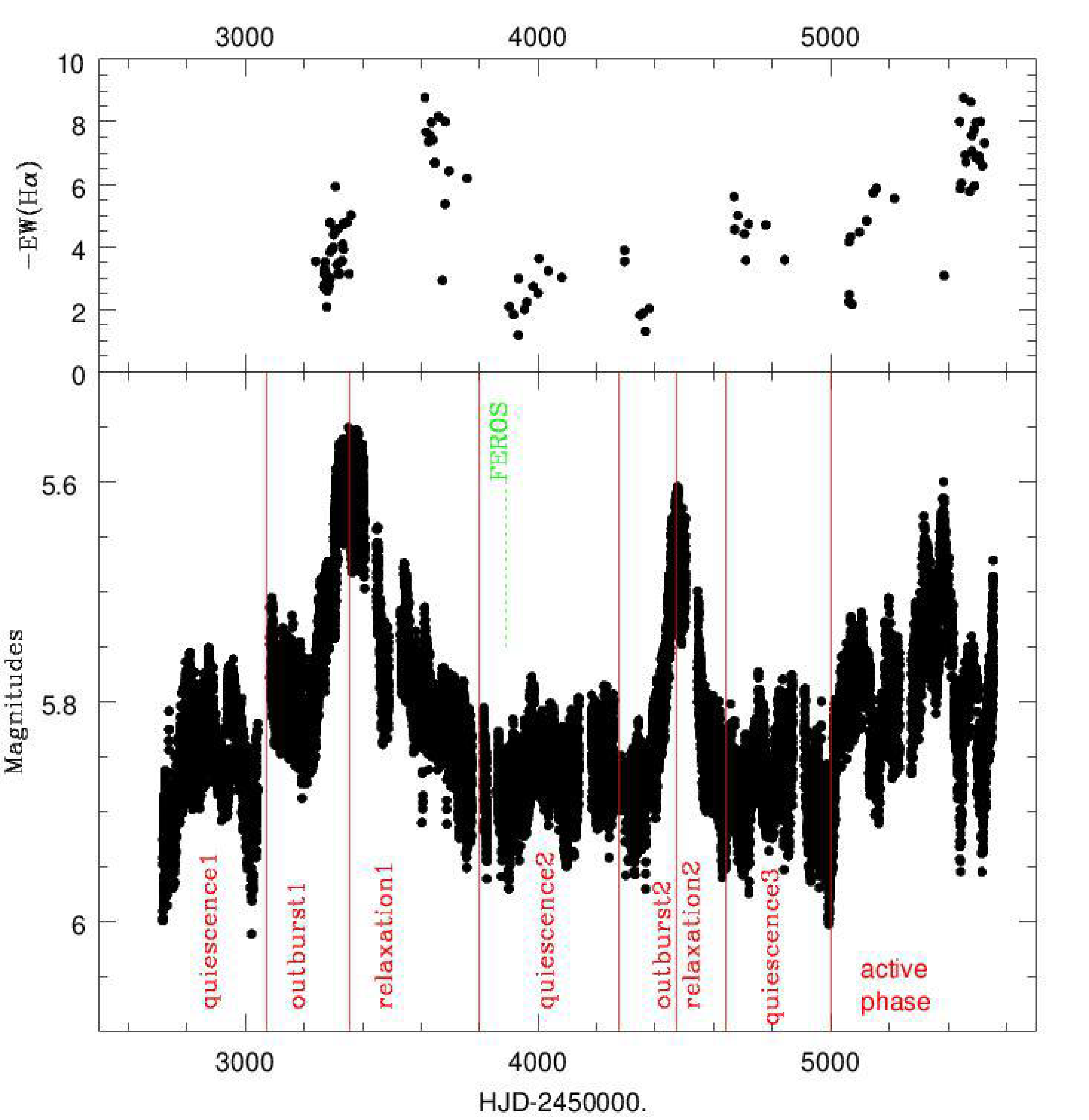}
  \end{center}
  \caption{Variation of the equivalent width of the H$\alpha$ line (which is in emission, hence $EW<0$) of \pa\ compared to the cleaned (but not detrended) {\it SMEI} light curve of the star. Temporal intervals discussed at the end of Sect. 3.1.1 are shown in red. }
\label{ha}
\end{figure}

Over the 8\,yr long campaign, \pa\ exhibited long-term variations. In the {\it SMEI} light curve, two brightenings particularly stand out because of their amplitude and duration: one around HJD$\sim$2\,453\,365 (2004 Dec. 25, $\Delta(mag)\sim0.25$\,mag and $\Delta(t)\sim650$\,d) and one around HJD$\sim$2\,454\,500 (2008 Feb. 3, $\Delta(mag)\sim0.2$\,mag and $\Delta(t)\sim300$\,d). Such variations are not uncommon for Be stars in general (or $\gamma$\,Cas analogs in particular). Figure \ref{ha} compares the light curve with the evolution of the H$\alpha$ line strength reported by \citet{zha13}. There is no obvious correlation between $EW(H\alpha)$ and the broad-band {\it SMEI} photometry, but the coverage of $EW$ measurements is scarce. A correlation with stronger H$\alpha$ emission when the star was brighter was reported for the 2013--2019 period ($\Delta(V)\sim0.4$\,mag while $EW$ changed from $\sim$0 to --25\,\AA, \citealt{naz19}), but the H$\alpha$ line dramatically changed in that interval, contrary to the relatively small variations that occurred during the {\it SMEI} observations. \pa\ therefore exhibits long-term photometric variations, possibly of diverse origins.

\begin{figure}
  \begin{center}
\includegraphics[width=8.5cm, bb = 3 3 680 730, clip]{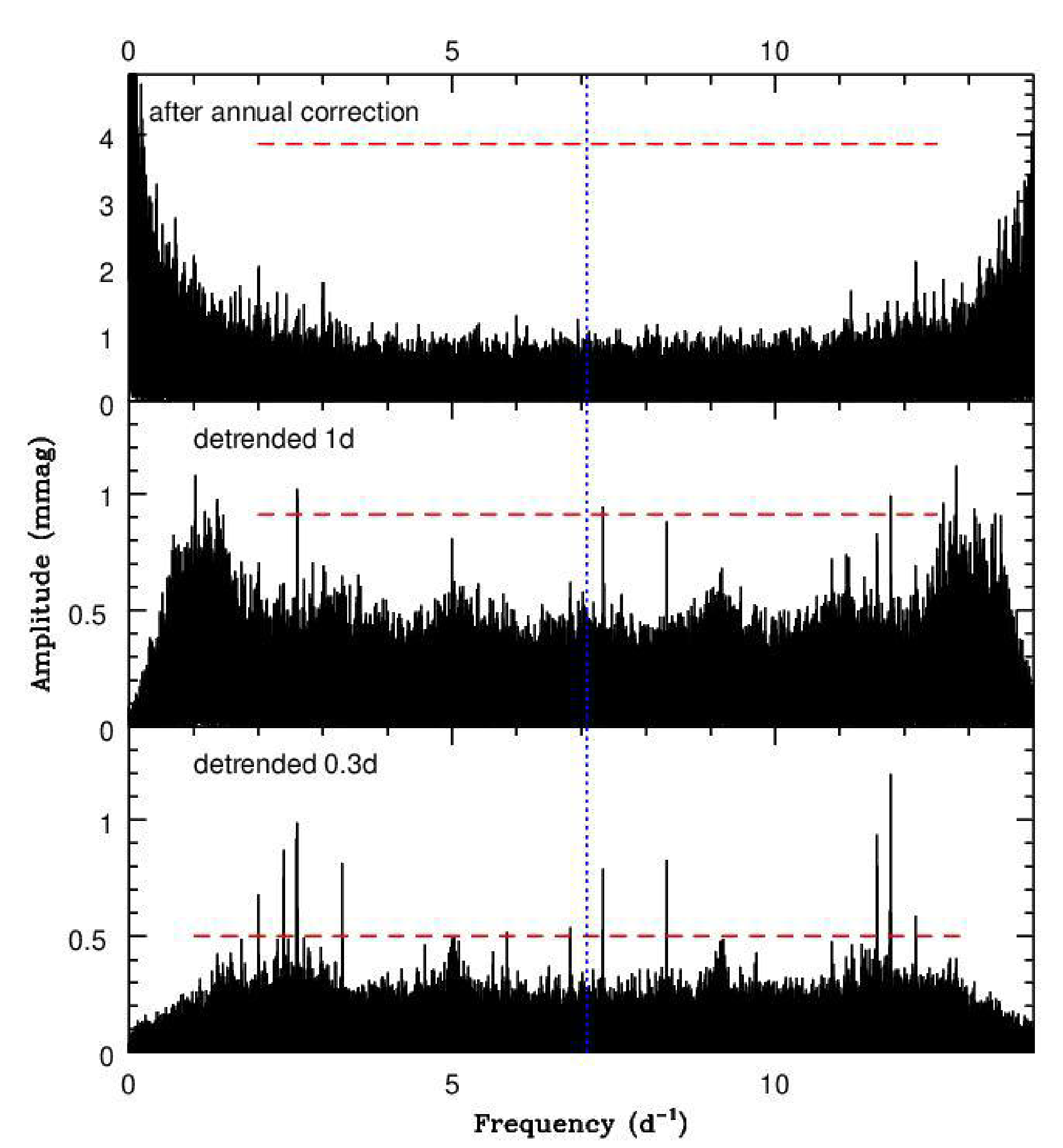}
\includegraphics[width=8.5cm, bb=19 300 591 693,clip]{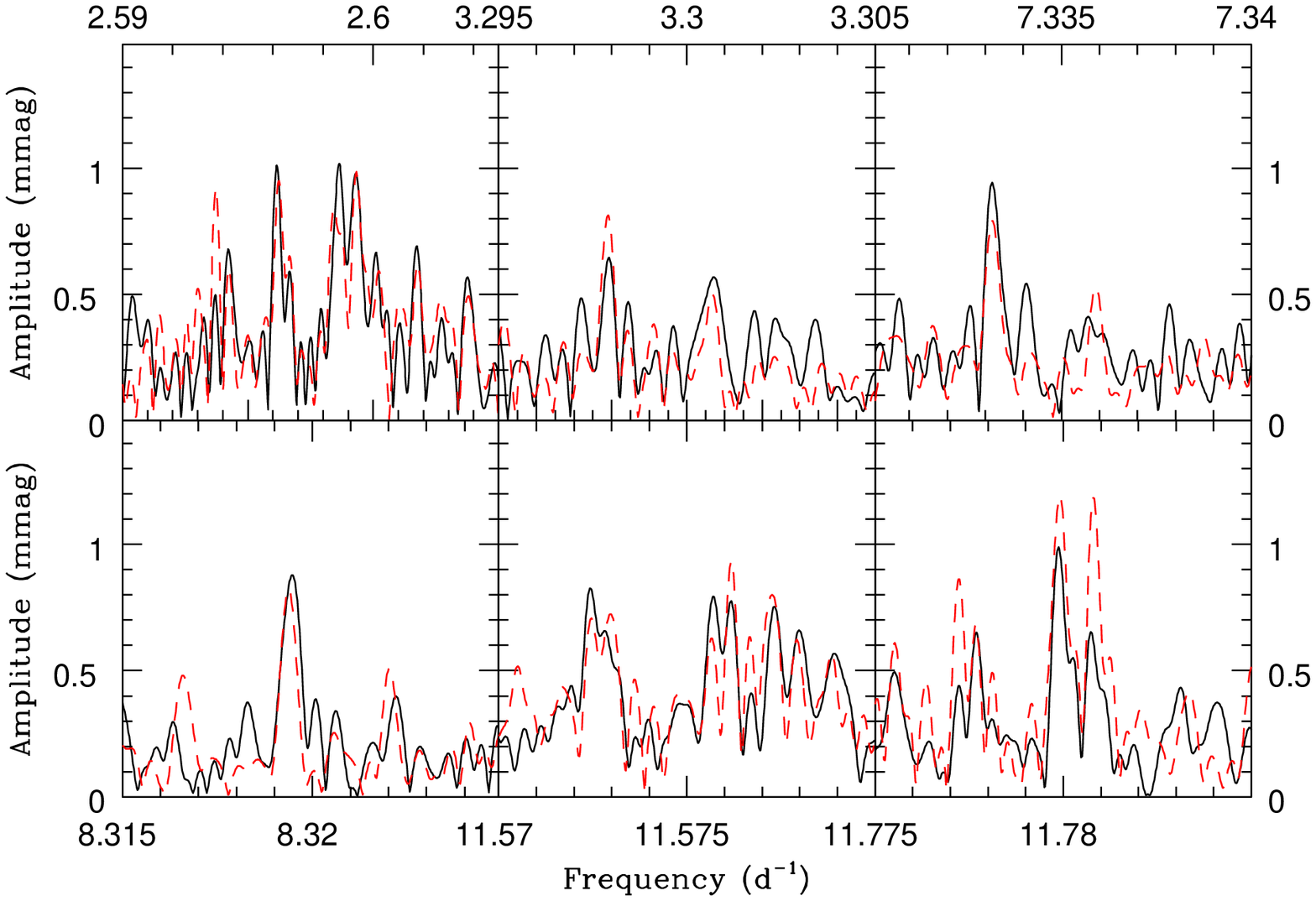}
  \end{center}
\caption{Frequency spectra derived by the modified Fourier algorithm for the {\it SMEI} light curves of \pa\ at various stages of processing. First panel from top - after correcting for yearly modulation and removing outliers, second panel from top - after those initial corrections and detrending with a 1\,d window, third panel from top - after correcting for yearly modulation, a stricter outlier rejection, and detrending with a 0.3\,d window. The blue vertical dotted line indicates the Nyquist frequency (half the inverse of the satellite orbital period) while the horizontal dashed red line provides the 1\% significance level (calculated using the formula of \citealt{mah11}). Bottom panels yield close-ups on the detected frequencies, to show their profile (the solid black line corresponds to results for the 1\,d detrended data, the dashed red line to the 0.3\,d detrending). }
\label{fourier}
\end{figure}

To find whether periodic variations are present, we applied to all cleaned {\it SMEI} light curves a Fourier algorithm adapted to datasets with uneven sampling\footnote{The Fourier method for period searches is equivalent to a least-squares fitting of a cosine curve to the observed data. In case of uneven sampling, however, the $\cos(\omega t_i)$ and $\sin(\omega t_i)$ terms are not independant anymore, rendering the calculation more complex (the full variance-covariance matrix must be considered). \citet{sca82} proposed such a calculation, but his method fits only a function $a \sin (\omega t)+ b \cos(\omega t)$ to the data as the mean of the data is subtracted in a preliminary step. However, this mean may not well represent the independent term and therefore, as the estimator of the mean is biased, his expression for the Fourier algorithm is necessarily also biased. The popular Scargle (or Lomb-Scargle) method is thus not fully robust, in particular "when the number of observations is small, the sampling is uneven, or the period of the sinusoid is comparable to or greater than the duration of the observations" \citep{cum99}. For this reason, \citet[see also \citealt{gos01}]{hmm} and \citet{zec09} introduced a modified Fourier algorithm which actually fits the variable component of the signal fully independently of the constant term thereby avoiding any bias. This is the method which is referred as "modified Fourier algorithm" in this paper.}. While the light curve sampling is not perfectly even, the gaps generally remain few in number and should not affect much the results (shown in Fig. \ref{fourier}). Indeed, very similar results are found when using the usual discrete Fourier transform. In this context, the binned analyses of variances (e.g. AOV, \citealt{sch89}) and conditional entropy \citep[see also \citealt{gra13}]{cin99,cin99b} were strongly affected by the satellite orbital period, leading to the presence of many subharmonics and aliases which renders them useless. Finally, since the data cover 2887.7\,d, the natural peak width is $3.5\times 10^{-4}$\,d$^{-1}$ in the periodogram. The precision on the frequency values are a fraction of this, $\sim5\times 10^{-5}$\,d$^{-1}$. 

\begin{figure}
  \begin{center}
\includegraphics[width=8cm, bb=3 3 415 740, clip]{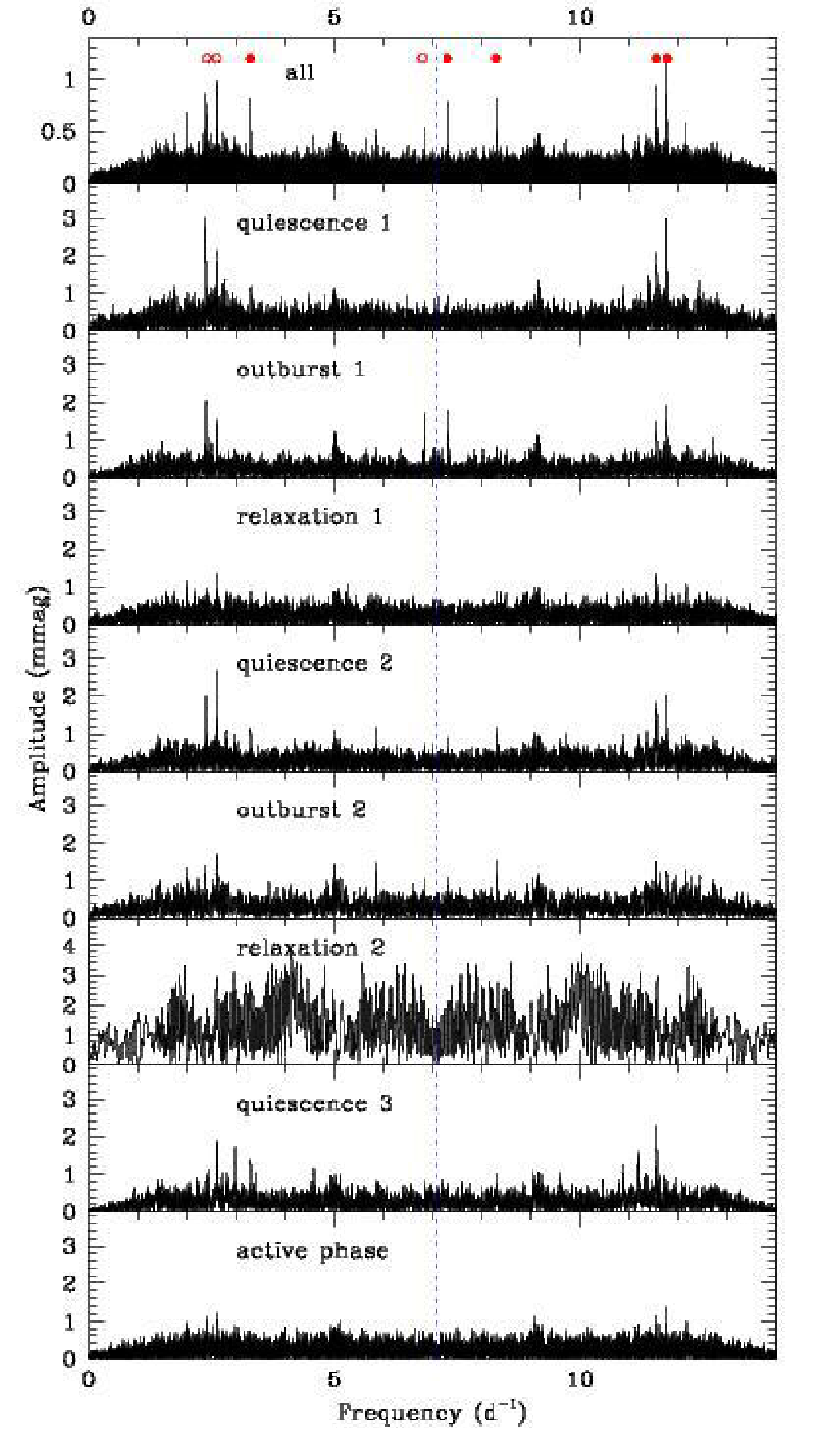}
\includegraphics[width=6cm, bb=35 200 555 690, clip]{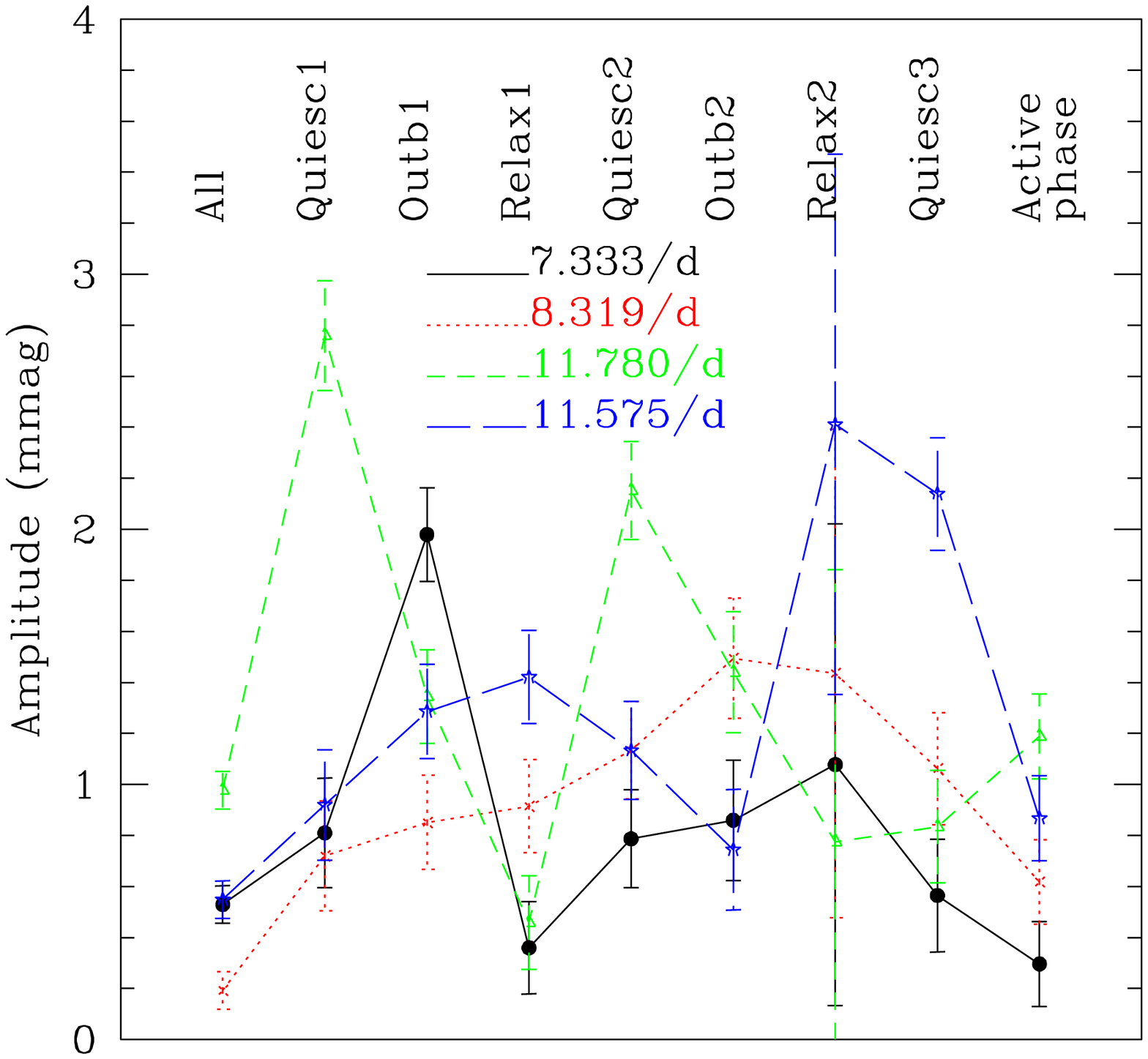}
  \end{center}
  \caption{{\it Top:} Appearance of the periodogram for different phases of activity in \pa\ (outburst, quiescence, and so on - see Fig. \ref{ha}). The data used for this analysis are those cleaned by the second method (i.e. with 0.3\,d detrending). Red dots in top panel indicate the astrophysical frequencies mentioned in the text, red circles are used for their aliases. The vertical blue dotted line is placed at the Nyquist frequency. {\it Bottom:} Amplitudes resulting from fitting a four-sine model to the photometric datasets cleaned by the second method, showing their variations as a function of the behaviour of \pa. }
\label{tempo2}
\end{figure}

The periodograms associated with the light curve cleaned but not detrended reveal strong signals at low frequencies, as could be expected in view of the presence of long-term variations. They may hide periodic short-term variations hence they were taken out by detrending. Indeed, the presence of several high-frequency signals is detected once this is done (Fig. \ref{fourier}). 

\begin{figure*}
  \begin{center}
\includegraphics[width=5.5cm, bb=0 110 390 730, clip, angle=90]{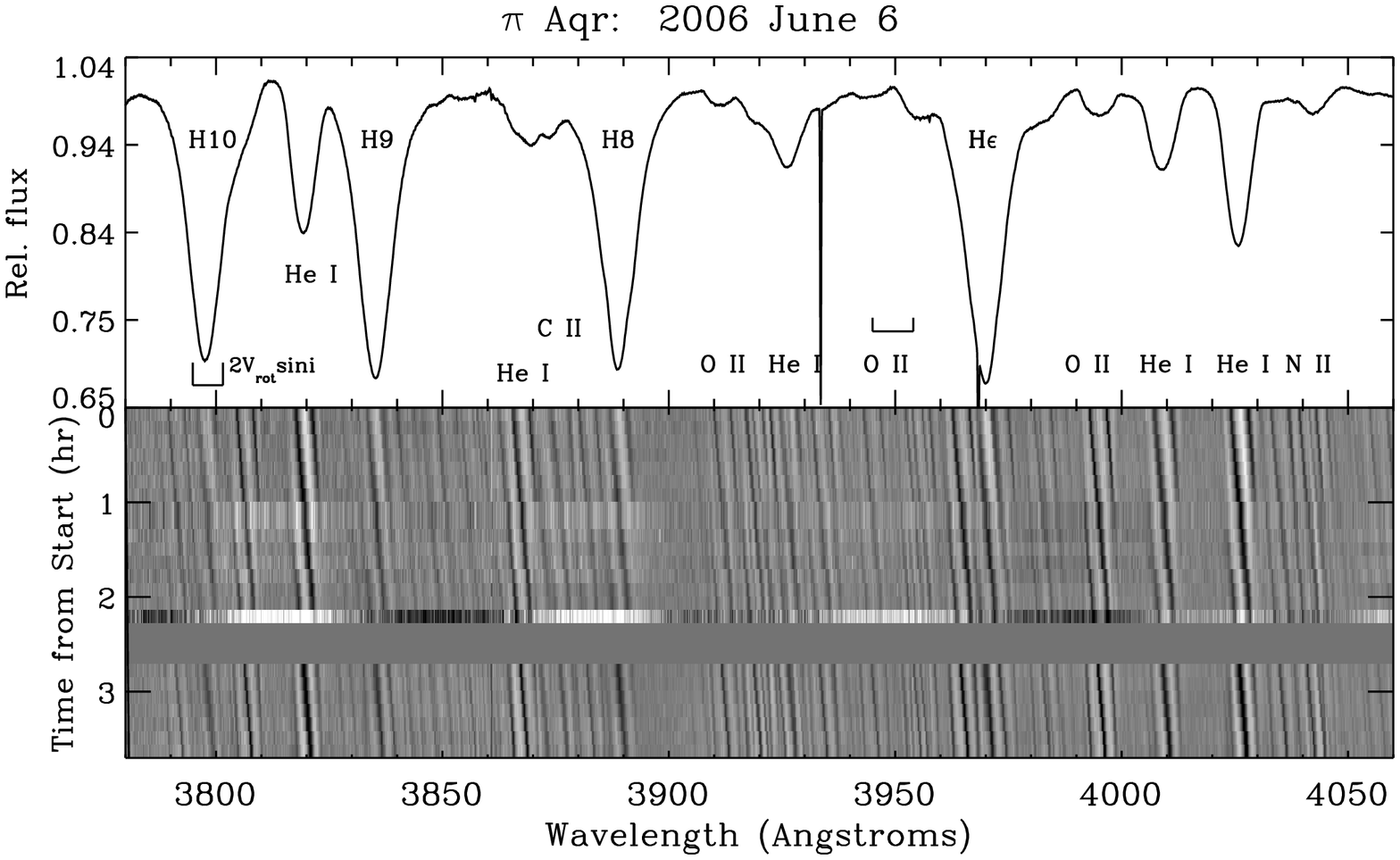}
\includegraphics[width=5.5cm, bb=0 110 390 730, clip, angle=90]{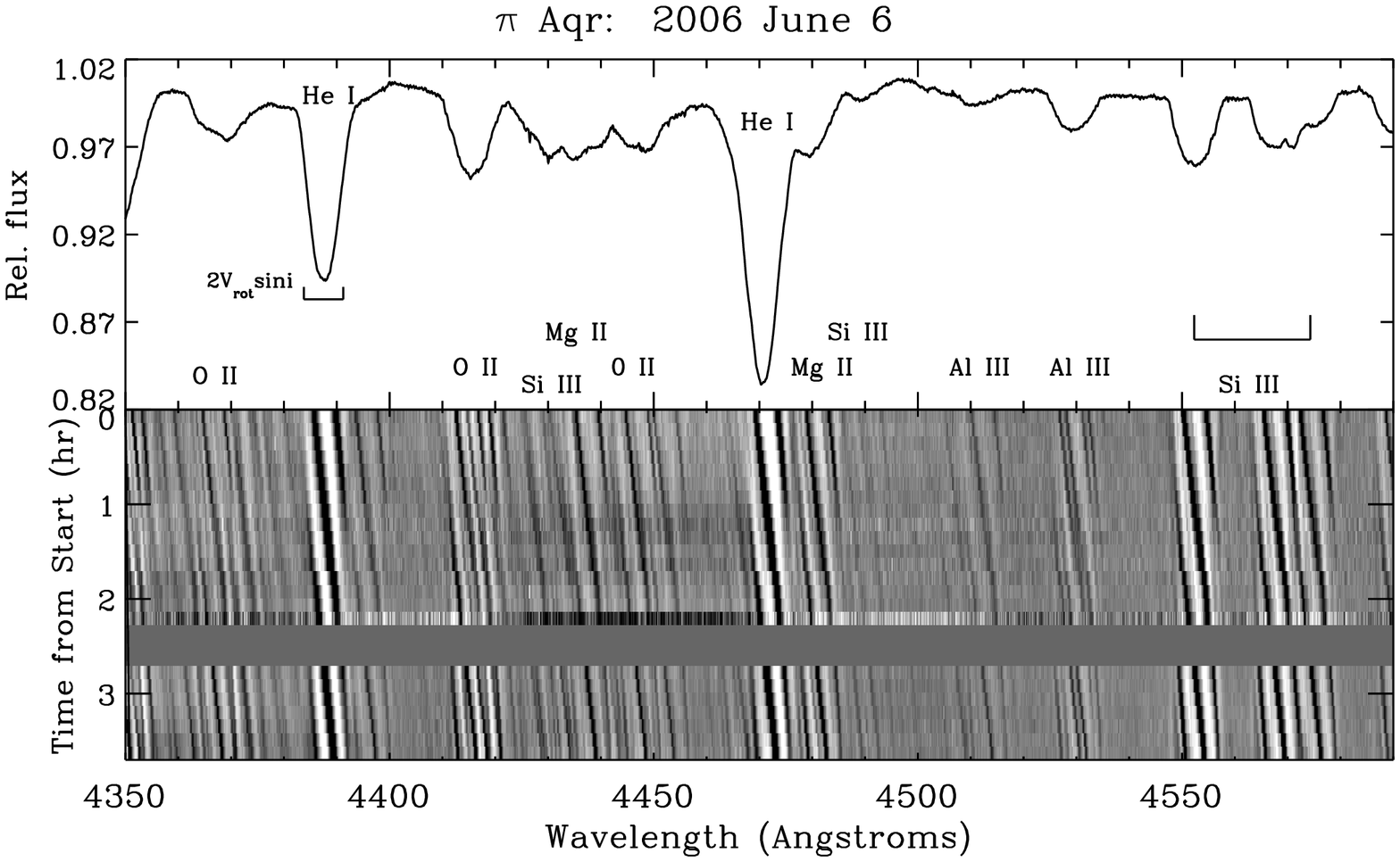}
\includegraphics[width=5.5cm, bb=0 110 390 730, clip, angle=90]{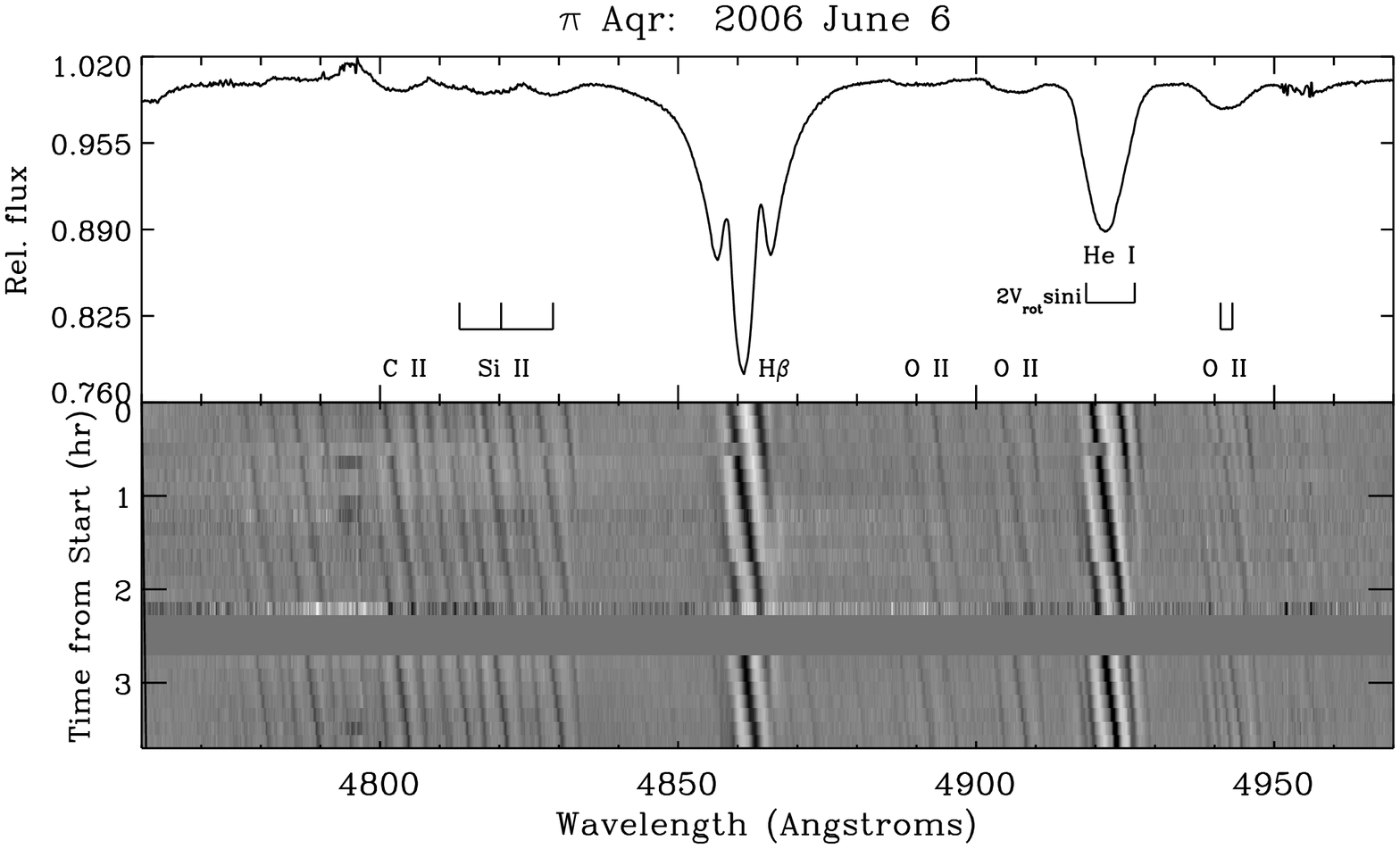}
\includegraphics[width=5.5cm, bb=0 110 390 730, clip, angle=90]{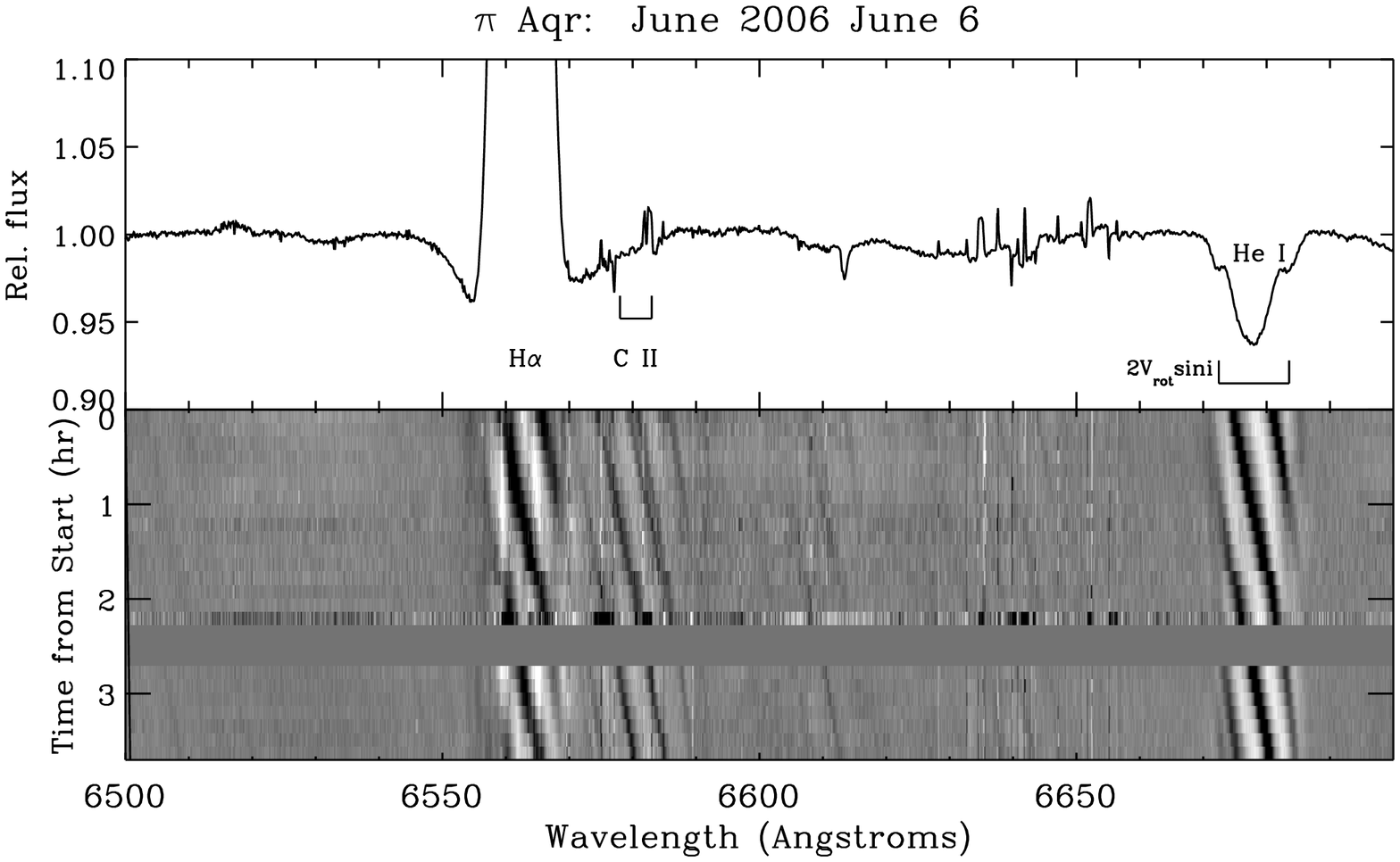}
  \end{center}
  \caption{Top panels display the mean FEROS spectrum of \pa, with prominent lines identified. Bottom panels provide greyscale difference spectra for 2006 June 06 (second FEROS night), with time running towards the bottom. A uniform dark grey row (at $\sim$2.6\,h) serves as a filler when no observation was taken.}
\label{graysc}
\end{figure*}

The main characteristic is a pair of frequencies near 11.8\,d$^{-1}$: 11.77988\,d$^{-1}$ and 11.78074\,d$^{-1}$ for the periodogram derived for data cleaned by the first method (with 1\,d detrending), or 11.77992\,d$^{-1}$ and 11.78080\,d$^{-1}$ for the second method (with 0.3\,d detrending). The small difference between frequency values found for the two cleaning methods are not significant considering the length of the dataset (see above). On the other hand, the frequency separation between the pair corresponds to $\sim$3.1\,yr, the interval between the two outbursts. It could be due to changes in amplitude of this signal (see more about this below) or instead be linked to an (unknown) instrumental effect. Note that the satellite orbited the Earth in 101.59\,min (averaged over the mission), which corresponds to a frequency of 14.175\,d$^{-1}$ hence a Nyquist frequency of 7.087\,d$^{-1}$. The aliases of a frequency $f_*$, given this satellite orbital characteristic, will be found near $f_{\rm orb}-f_*\sim$2.4\,d$^{-1}$ (they are especially visible for data cleaned by the second method).

At frequencies of 11.5724--11.5773\,d$^{-1}$ a second group of peaks appears. This group displays an alias at 2.5937--2.5993\,d$^{-1}$.

Two additional frequencies are also clearly seen: isolated peaks at 7.33318\,d$^{-1}$ (same value for data cleaned by both methods) and at 8.31948\,d$^{-1}$ (or 8.31938\,d$^{-1}$ for the second method).  The alias of the 7.3\,d$^{-1}$ peak is found near 6.8\,d$^{-1}$.

One frequency, 3.29790\,d$^{-1}$, is clearly seen only for the data cleaned by the second method. Some signal exists at that frequency in the data from the first method, but it is not significant. The reason for this is unknown as the two processing methods generally yield very similar results (Fig. \ref{fourier}).

Some additional frequencies, at 2.0\,d$^{-1}$ or 5.0\,d$^{-1}$, are visible but they are likely spurious. The analysis of {\it SMEI} data often yields periods at integer frequencies due to the sun-synchronous orbit hence any real stellar signal would be lost at these frequencies \citep{gos11}.

Using Eq. (2) of \citet{mah11}, we derive the significance levels of these frequencies (see also Fig. \ref{fourier}). For the data cleaned by the first method, significance levels reach 0.03\% for 2.6\,d$^{-1}$, 0.4\% for 7.3\,d$^{-1}$, 3\% for 8.3\,d$^{-1}$, 13\% for 11.6\,d$^{-1}$, and 0.08\% for 11.8\,d$^{-1}$.  For the data cleaned by the second method, they are even better (Fig. \ref{fourier}). We thus consider them all significant. 

Next, we study how the frequency content varies with time. We define several intervals depending on the long-term behaviour (Fig. \ref{ha}): quiescence, rise of outburst, relaxation after outburst, and ``active phase'' (when there is no clear quiescence and small brightenings are spotted). The resulting periodograms are shown at the top of Fig. \ref{tempo2}. In parallel, a 4-sine model with frequencies fixed at 7.333, 8.319, 11.78, and 11.575\,d$^{-1}$ was fitted to the light curves: the resulting amplitudes evolve in a similar way as shown by periodograms (see bottom of Fig. \ref{tempo2}) despite the simplicity of such a model and the complexity of the periodogram peaks (see bottom of Fig. \ref{fourier}). Clearly, the amplitudes of the different signals vary with time. The 11.8\,d$^{-1}$ frequency appears stronger during the first ``outburst'' and the first two ``quiescence'' intervals. The 3.3\,d$^{-1}$ and 11.6\,d$^{-1}$ signals are stronger during all ``quiescence'' intervals. During the first ``quiescence'' and first ``outburst'', the 11.8\,d$^{-1}$ signal dominates the neighbouring 11.6\,d$^{-1}$ signal, while the situation is reversed during the third ``quiescence'' and their amplitudes are similar otherwise. The amplitudes of both 11.6\,d$^{-1}$ and 11.8\,d$^{-1}$ signals are much reduced in the first ``relaxation'' and ``active'' phases. In parallel, the 7.3\,d$^{-1}$ signal appears strong during the two ``outburst'' phases, especially during the first one, whereas the 8.3\,d$^{-1}$ signal reaches its maximum amplitude during the second ``outburst''. Both frequencies are barely visible during the second ``quiescence'' phase and appear weak during the other phases. More generally, during ``relaxation'' and ``active'' phases, the periodograms appear rather flat, with only weak signs of those frequencies. This is not due to noise, except for the second relaxation episode (because of its smaller number of data points): the periodic signals truly are weaker at those times. 

\begin{figure}
  \begin{center}
\includegraphics[width=7cm, bb= 110 167 535 717, clip]{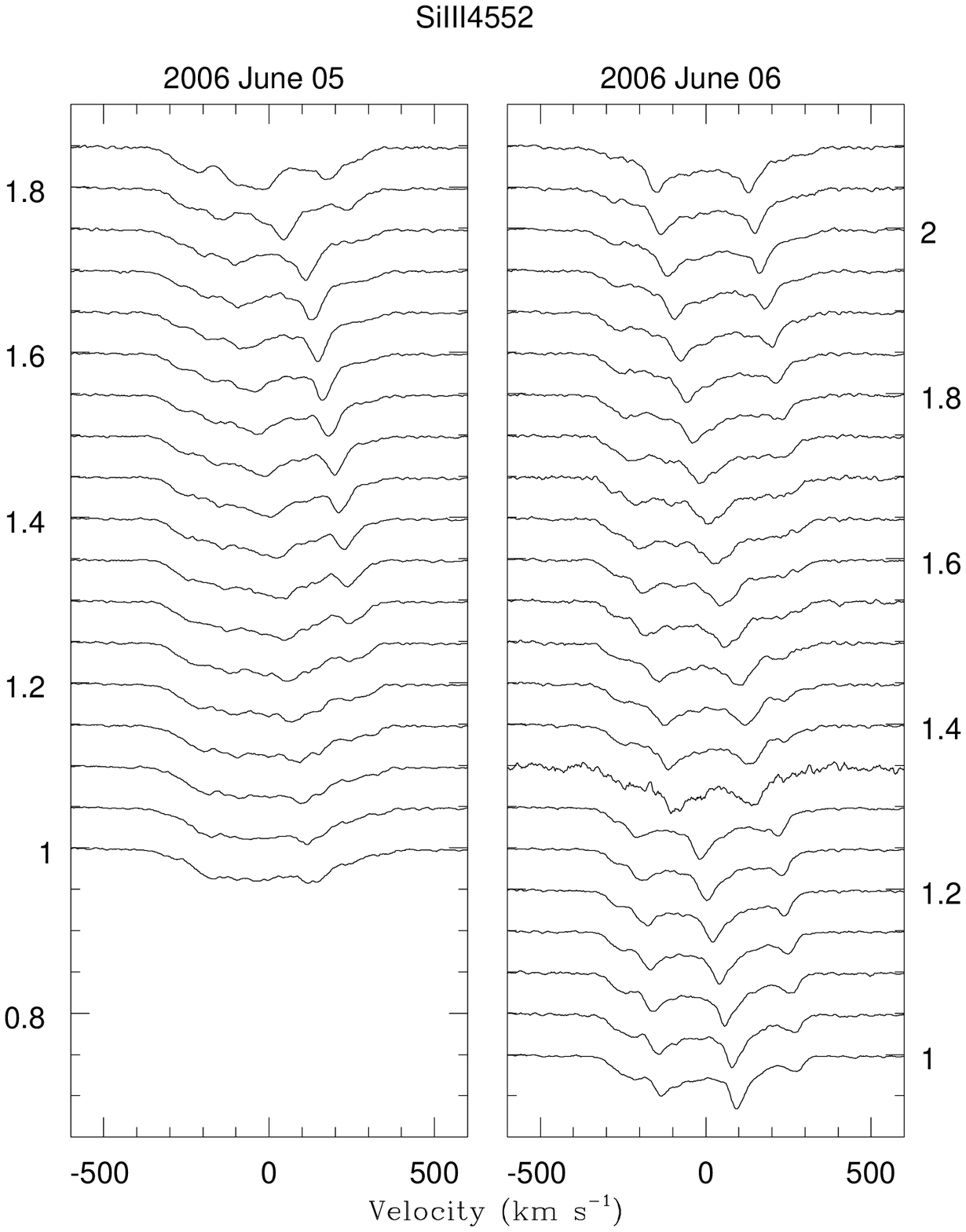}
  \end{center}
  \caption{Evolution of the line profile of Si\,{\sc iii}\,$\lambda$4552\,\AA\ during the FEROS monitoring of \pa. The profiles recorded on the first night are shown on the left and those of the second night on the right, with time increasing towards the bottom and an arbitrary vertical shift of 0.05 between consecutive spectra to facilitate comparison. }
\label{prof}
\end{figure}

\subsubsection{Spectroscopy}
Variations in the spectrum of \pa\ have been detected for nearly a century, as its disk may completely dissipate or present various levels of emission \citep[e.g.][]{mcl62,bjo02}. In June 2006, the disk activity of \pa\ was moderate, with an overall equivalent width ($EW$) for its H$\alpha$ line of about --2.7\,\AA. For comparison, a recent monitoring indicated $EW\sim-23$\,\AA\ in 2018 \citep{naz19}. In the {\it SMEI} light curve (Fig. \ref{ha}), the FEROS observations correspond to the beginning of the second quiescence.

As a consequence, the 2006 optical spectra mostly show absorption lines from H\,{\sc i}, He\,{\sc i}, C\,{\sc ii}, N\,{\sc ii}, O\,{\sc ii},  Mg\,{\sc ii}, Si\,{\sc iii}, and Fe\,{\sc iii} (Fig. \ref{graysc}). Emission is limited to the H$\alpha$ line, though some residual emission obviously pollutes the other Balmer lines, e.g. H$\beta$. Compared to the 2018 monitoring, we note the absence of the numerous Fe\,{\sc ii} emissions, but they had previously been found to disappear when disk activity was low or moderate \citep{mcl62}.

\begin{figure}
  \begin{center}
\includegraphics[width=8.5cm, bb=40 160 490 720, clip]{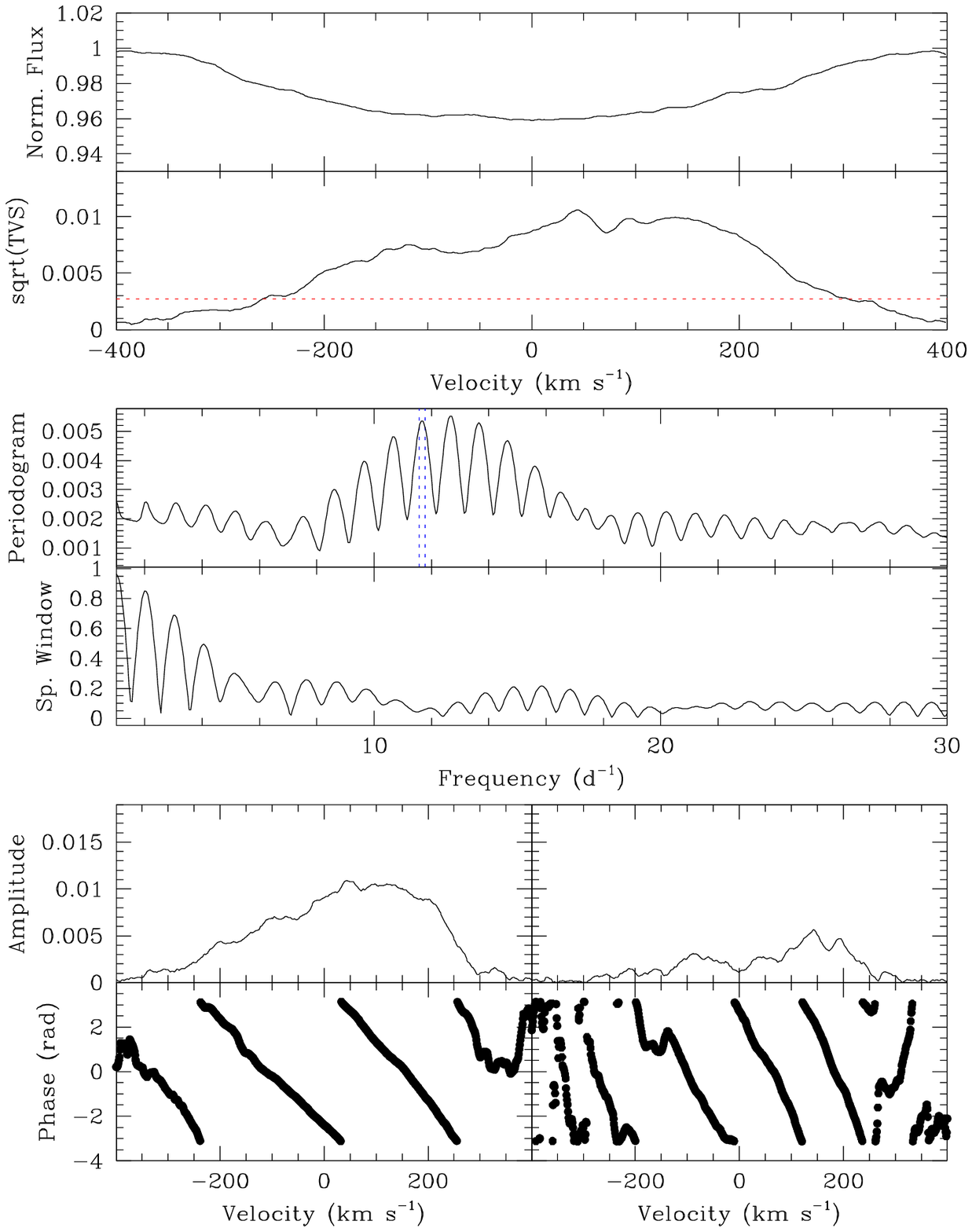}
  \end{center}
  \caption{Variability properties of Si\,{\sc iii}\,$\lambda$4552\,\AA\ in the FEROS data of \pa. {\it Top panels:} mean line profile and TVS (with the 1\% significance level in red). {\it Middle panels:} periodogram calculated on the line profile (in the same velocity range as shown above and with the 11.6\,d and 11.8\,d  frequencies shown by dotted blue lines) and spectral window associated with the sampling. {\it Bottom panels:} amplitude and phase of the best-fit double sine wave ($f+2\times f$ with $f=11.78$\,d$^{-1}$; left panels correspond to the $f$ component and the right ones to the $2f$ component). }
\label{pgram}
\end{figure}

Focusing on isolated strong lines (He\,{\sc i}\,$\lambda\lambda$4026,\,4388, 4471,\,4921,\,6678\,\AA, Si\,{\sc iii}\,$\lambda\lambda$4552,\,4567,\,4574\,\AA, H$\alpha$, H$\beta$) immediately reveals small-scale variations of the line profile (Fig. \ref{prof}), with several fast, blue-to-red moving absorption features superimposed on the photospheric absorption. These so-called ``migrating subfeatures'' travel through the line profiles at a rate of $\sim$3000\,\kms\,d$^{-1}$ (or $\sim$125\,\kms\,hr$^{-1}$). They are particularly strong and narrow for He\,{\sc i}\,$\lambda$6678\,\AA\ and Si\,{\sc iii}\,$\lambda\lambda$4552--74\,\AA. Comparing He\,{\sc i} lines with one another, the sharpness of these features clearly increases with increasing wavelength, especially for singlets (He\,{\sc i}\,$\lambda\lambda$4388,\,4921,\,6678\,\AA). They are at first more difficult to see on the H\,{\sc i} lines, but subtracting the mean profile clearly reveals the presence of the same variability pattern (Fig. \ref{graysc}). Interestingly (see Sect. 4.1), the greyscale difference spectra reveal the same pattern redwards of H$\alpha$, at the position of the C\,{\sc ii}\,$\lambda\lambda$6578,\,6583\,\AA\ doublet despite the fact that these lines are not detectable in the mean spectrum. 

\begin{figure*}
  \begin{center}
\includegraphics[width=6.3cm, bb=0 110 390 730, clip, angle=90]{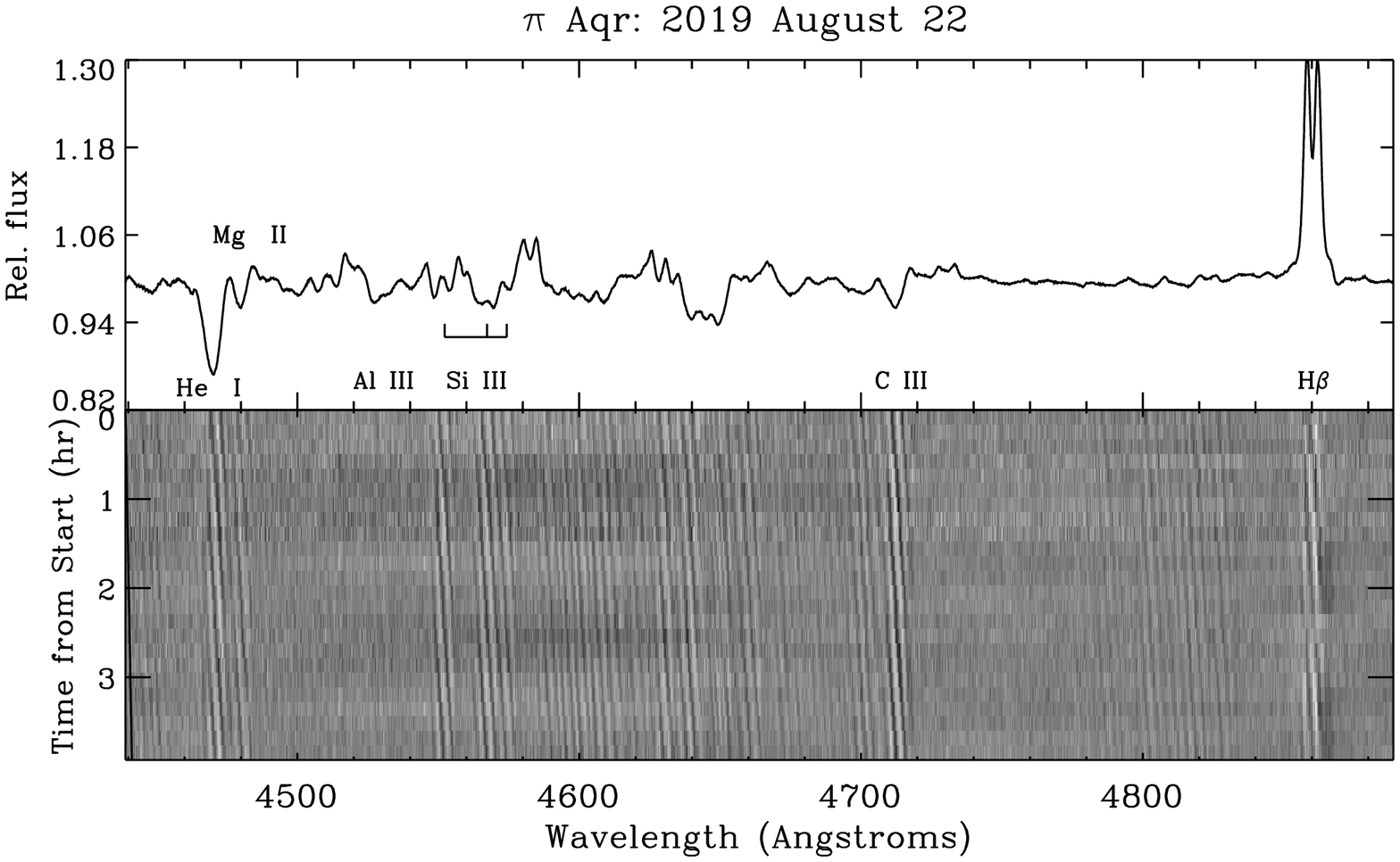}
\includegraphics[width=7.5cm, bb=40 160 490 720, clip]{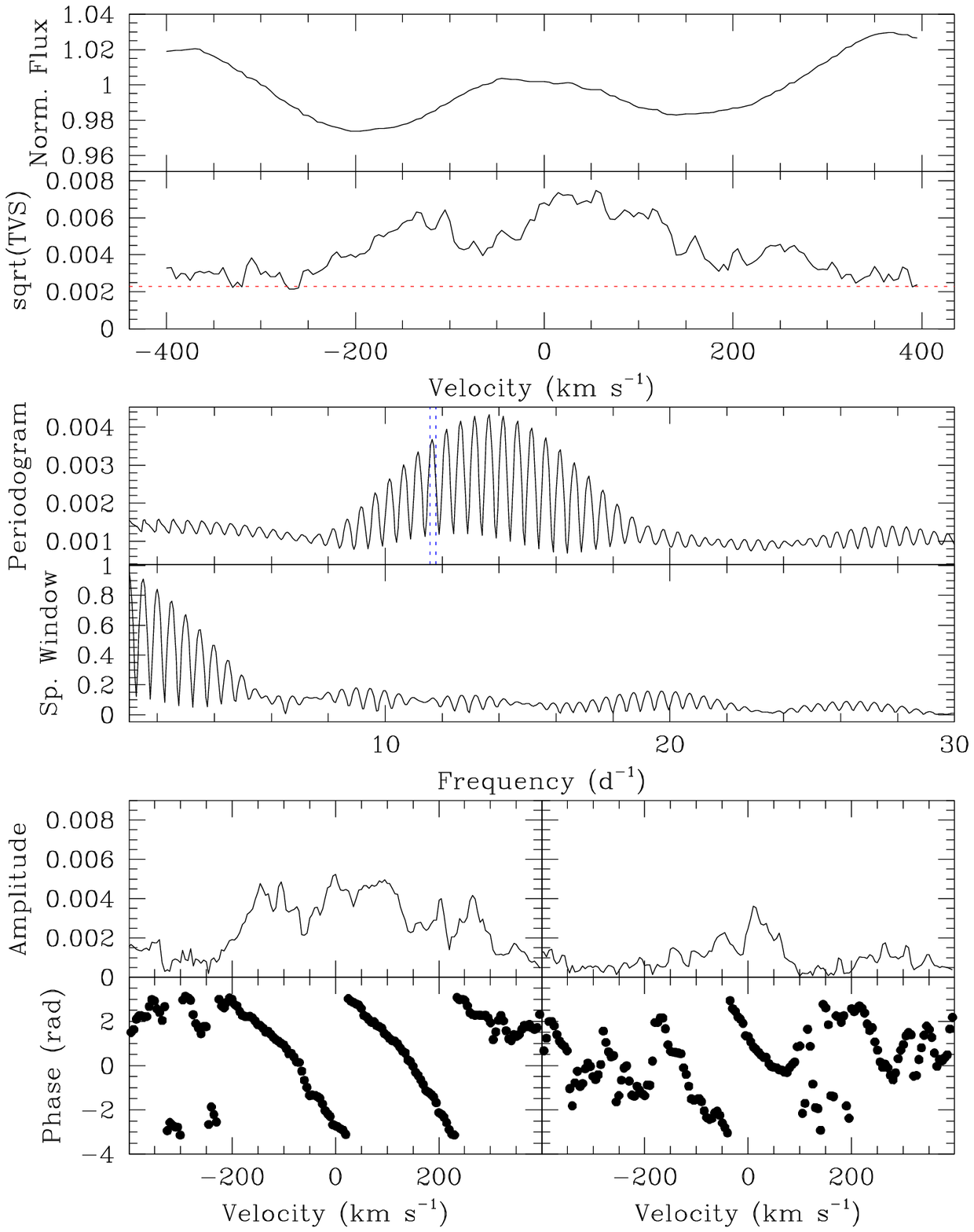}
  \end{center}
  \caption{Same as Figs. \ref{graysc} and \ref{pgram} for the OHP observing run. {\it Left panels:} Mean spectrum and greyscale difference spectra for the first observing night, after a simple mean was taken out from the line profiles. Time increases towards the bottom. {\it Right panels:} Average line profile of Si\,{\sc iii}\,$\lambda$4552\,\AA\ and associated TVS (with the 1\% significance level in red), periodogram of that line in the same velocity range and spectral window associated with the sampling, amplitude and phase of the best-fit double ($f+2\times f$ with $f$=11.78\,d$^{-1}$) sine wave. The results for the fundamental frequency are shown at left and those for the second harmonic at right. }
\label{ohp}
\end{figure*}

\begin{table*}
\centering
\caption{Average line moments and their dispersion for the FEROS monitoring of \pa. An asterisk indicates interstellar lines.}
\label{mom}
\setlength{\tabcolsep}{3.3pt}
\begin{tabular}{lccccc}
\hline\hline
Line & $\lambda_0$ (\AA) & $\Delta(V)$ (km\,s$^{-1}$) & $EW$ (\AA) & $M_1$ (km\,s$^{-1}$) & width (km\,s$^{-1}$) \\
\hline
H10                          & 3797.898 &--400. 400.  & 2.252$\pm$0.008 & --8.1$\pm$0.9 & 202.6$\pm$0.5 \\
H9                           & 3835.384 &--400. 400.  & 2.421$\pm$0.012 & --6.4$\pm$0.7 & 202.4$\pm$0.4 \\
H8                           & 3889.049 &--400. 400.  & 2.34$\pm$0.03   & --7.7$\pm$0.6 & 202.6$\pm$0.6 \\
He\,{\sc i}\,$\lambda$3819   & 3819.757 &--400. 400.  & 0.90$\pm$0.03   &--16.7$\pm$1.0 & 163.9$\pm$3.2 \\
He\,{\sc i}\,$\lambda$3927   & 3926.544 &--400. 400.  & 0.598$\pm$0.006 &--28.2$\pm$1.6 & 195.7$\pm$1.3 \\
Ca\,{\sc ii}\,$\lambda$3933$^*$ & 3933.663 &--25.   5.   &0.0922$\pm$0.0012&--9.74$\pm$0.04&  7.09$\pm$0.03\\
He\,{\sc i}\,$\lambda$4009   & 4009.256 &--400. 400.  & 0.529$\pm$0.006 &--26.6$\pm$2.1 & 162.1$\pm$1.3 \\
He\,{\sc i}\,$\lambda$4026   & 4026.357 &--400. 400.  & 1.042$\pm$0.011 &--38.3$\pm$1.3 & 165.2$\pm$1.3 \\
He\,{\sc i}\,$\lambda$4388   & 4387.929 &--400. 400.  & 0.710$\pm$0.005 &--14.9$\pm$1.7 & 162.4$\pm$1.3 \\
He\,{\sc i}\,$\lambda$4471   & 4471.512 &--400. 400.  & 1.239$\pm$0.010 &--39.1$\pm$1.8 & 183.7$\pm$0.9 \\
Mg\,{\sc ii}\,$\lambda$4481  & 4481.228 &--250. 250.  & 0.192$\pm$0.005 &--48.8$\pm$3.4 & 126.9$\pm$2.1 \\
Si\,{\sc iii}\,$\lambda$4552 & 4552.622 &--400. 400.  & 0.296$\pm$0.006 & --6.3$\pm$4.2 & 162.8$\pm$2.6 \\
H$\beta$                     & 4861.325 &--540. 540.  & 2.071$\pm$0.011 &--17.6$\pm$2.3 & 263.2$\pm$0.6 \\
He\,{\sc i}\,$\lambda$4921   & 4921.931 &--400. 400.  & 0.841$\pm$0.004 &--10.0$\pm$2.0 & 165.9$\pm$1.3 \\
H$\alpha$                    & 6562.85  &--540. 540.  & -2.73$\pm$0.03  & --6.7$\pm$3.1 &  89.2$\pm$3.9 \\
DIB\,$\lambda$6613$^*$       & 6613.62  &--50.  60.   & 0.035$\pm$0.003 & --0.6$\pm$1.7 &  26.2$\pm$1.2 \\
He\,{\sc i}\,$\lambda$6678   & 6678.151 &--400. 400.  & 0.498$\pm$0.014 &--11.3$\pm$7.4 & 145.6$\pm$4.7 \\
\hline      
\end{tabular}
\end{table*}

We calculated the moments of several lines as in \citet{naz19}\footnote{They are defined as $M_0=\sum (F_i-1)$, $M_1=\sum (F_i-1)\times v_i /\sum (F_i-1)$, and $M_2=\sum (F_i-1)\times (v_i-M_1)^2 /\sum (F_i-1)$ where $v_i$ is the radial velocity and $F_i$ the normalized flux. $M_0$ provides the equivalent width after multiplication by minus the wavelength step (to keep positive values for absorption lines), $M_1$ is the line centroid (or radial velocity), and the squared root of $M_2$ yields the line width.}. The averages and dispersions ($=\sqrt{\sum{(M_i-mean)^2}/(N-1)}$) of these moments were computed and are provided in Table \ref{mom}. Unsurprisingly, the largest dispersions are found for He\,{\sc i}\,$\lambda$6678\,\AA\ and Si\,{\sc iii}\,$\lambda$4552\,\AA, whatever the moment considered. The moment dispersions for the other lines depend on the moment under consideration, however. Based on the Ca\,{\sc ii} interstellar lines, equivalent widths of strong lines can be determined with a precision of 0.01\AA\ on these FEROS spectra. The stellar lines (except the two cases just mentioned) display similar dispersions, showing that the line profile variability does not much impact the equivalent width derivation. A similar situation occurs for line widths but the impact appears much more severe on line centroids. Peak-to-peak excursions up to 10\,km\,s$^{-1}$ can be found for stellar lines. This could easily explain the scatter observed around the orbital curve \citep{bjo02,naz19}. 

\begin{figure}
  \begin{center}
\includegraphics[width=8.cm,bb= 3 3 730 710,clip]{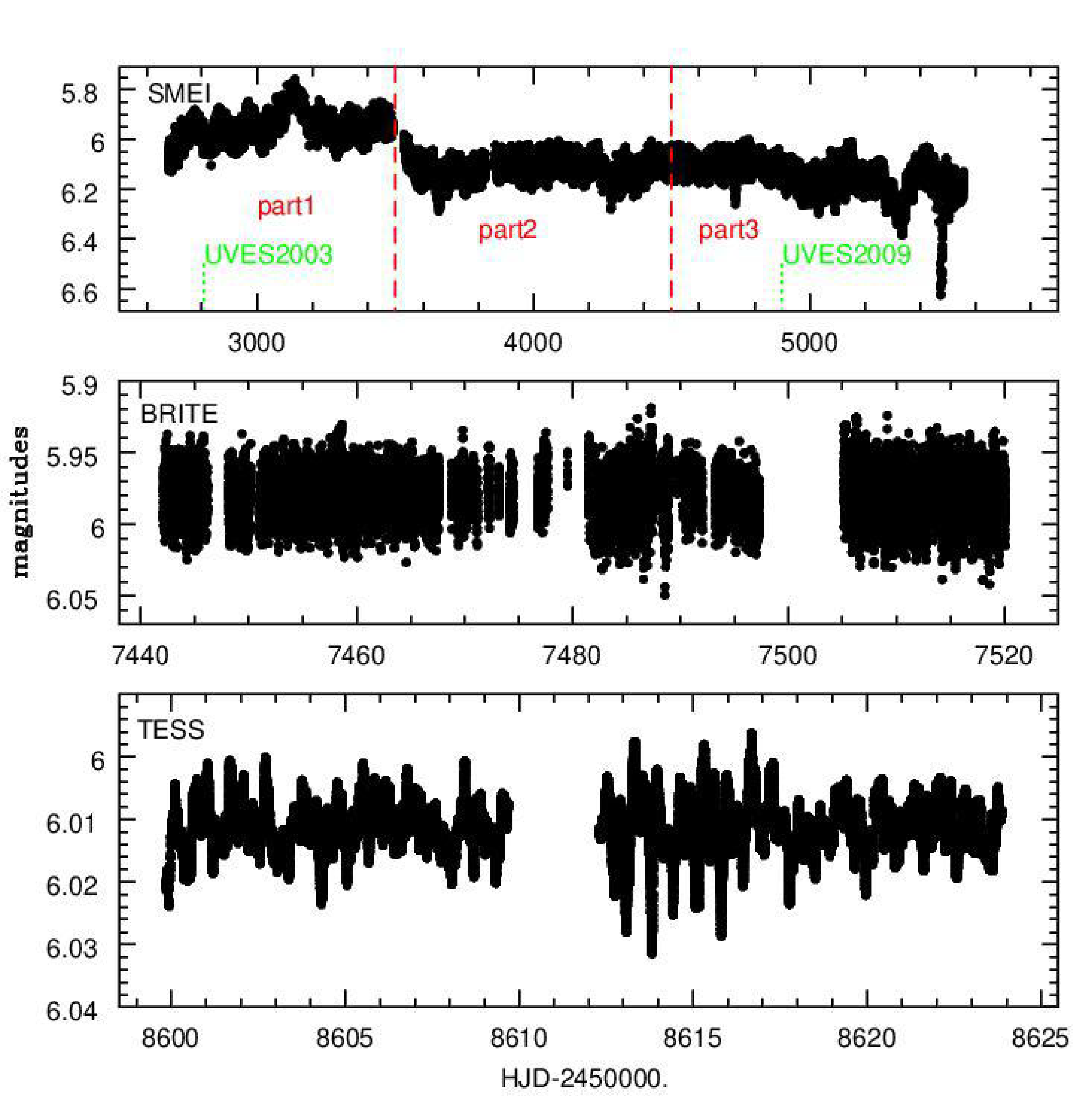}
  \end{center}
  \caption{Space-based photometric campaigns of \bz, with the limits of the temporal intervals discussed at the end of Sect. 3.2.1 marked by red dashed lines and with the time of two UVES monitorings indicated by green dotted lines (no photometric data is avalable at the time of the third one). }
\label{lcbz1}
\end{figure}

\begin{figure*}
  \begin{center}
\includegraphics[width=5.6cm,bb=30 150 560 720, clip]{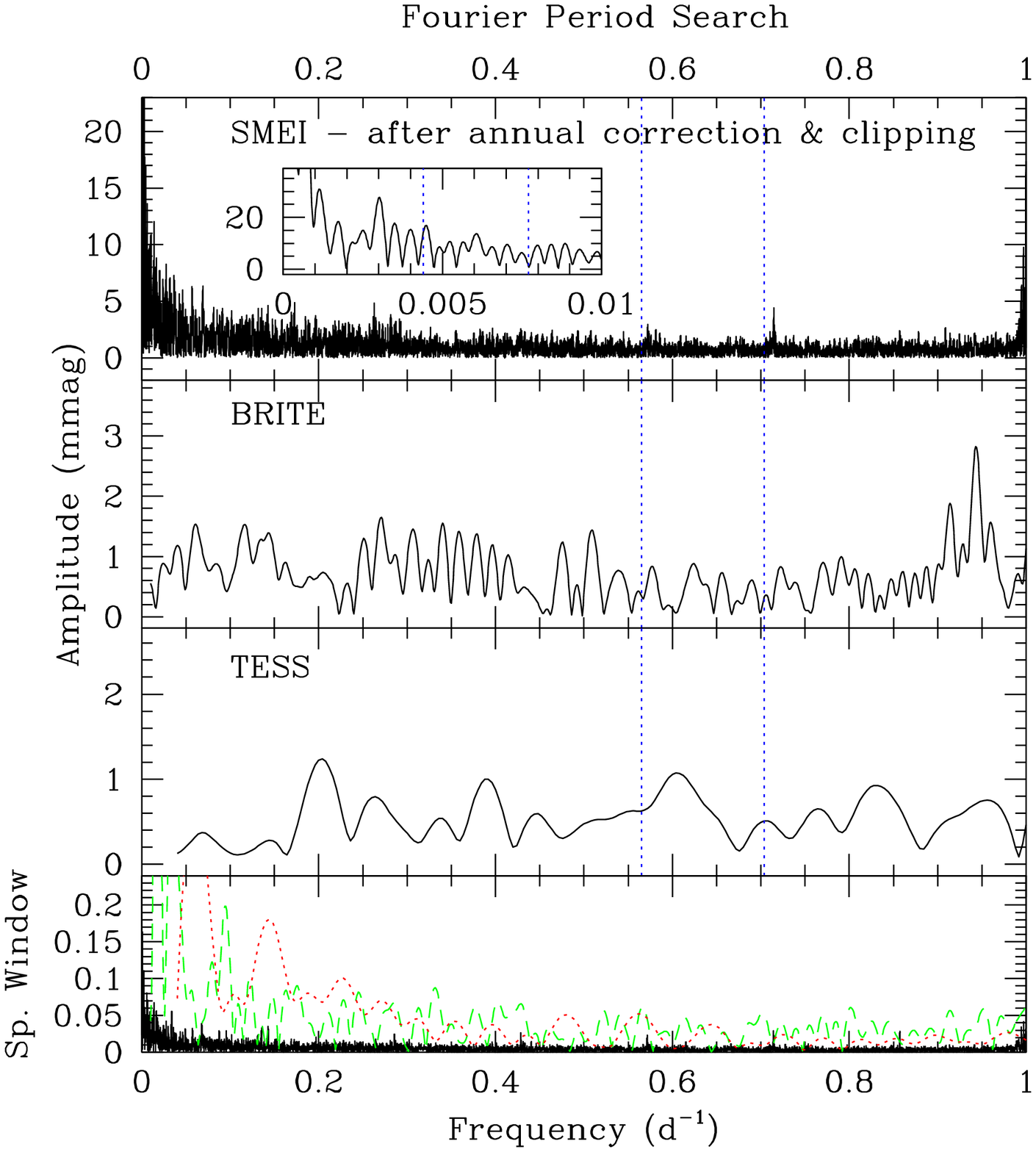}
\includegraphics[width=5.6cm,bb=3 3 675 755, clip]{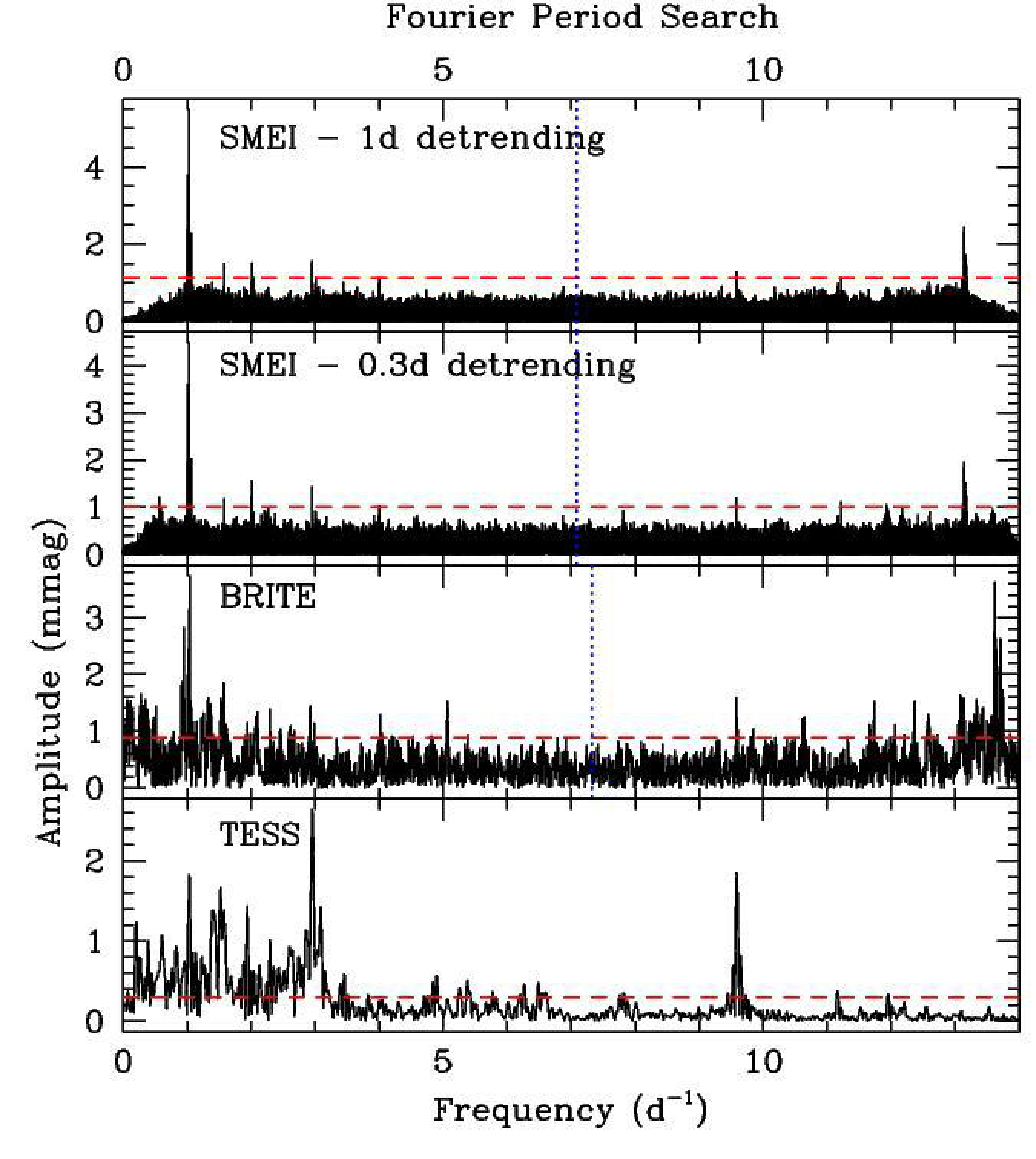}
\includegraphics[width=5.6cm,bb=50 150 560 720, clip]{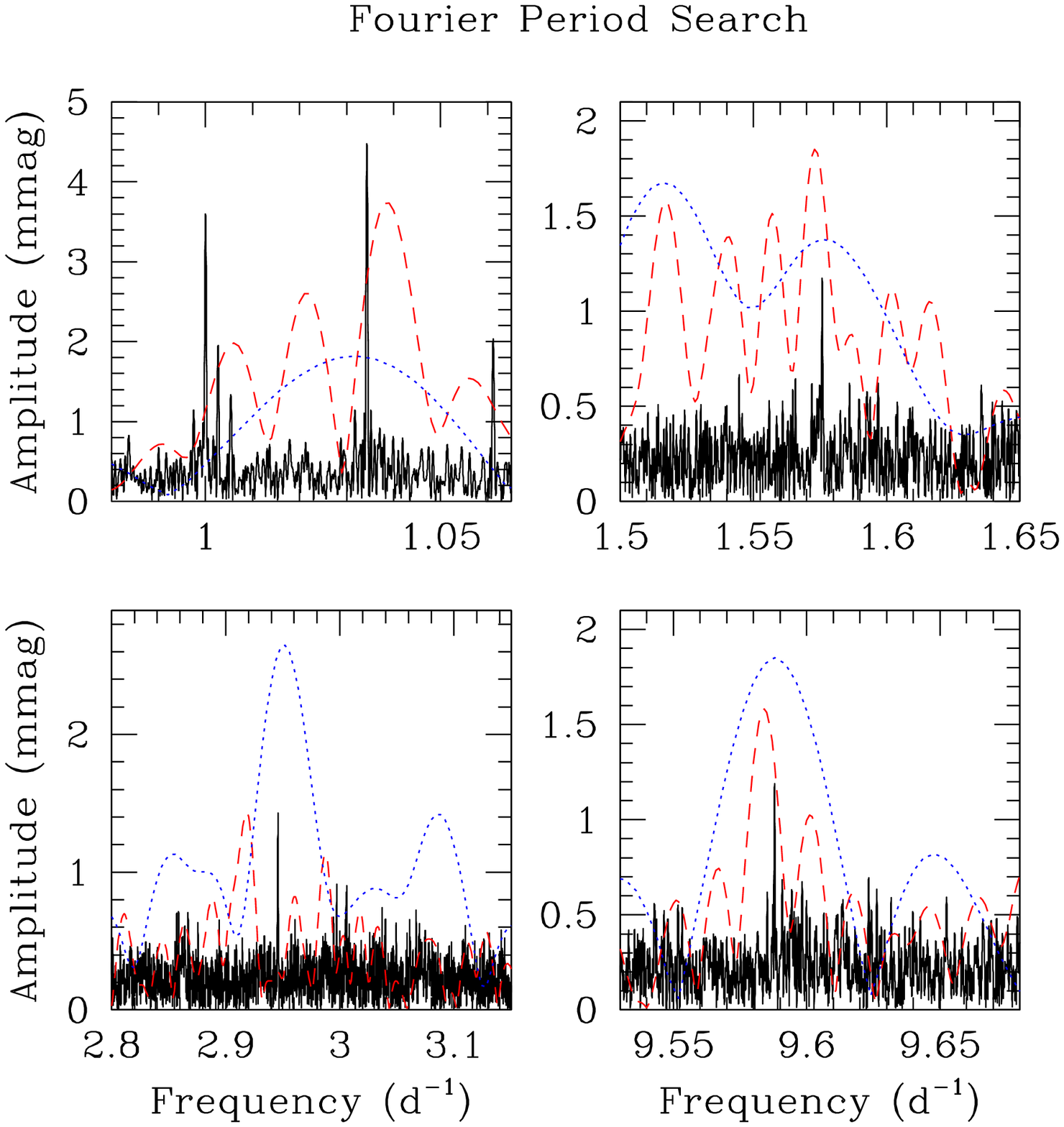}
  \end{center}
  \caption{Period search results for light curves of \bz. {\it Left:} Frequency spectra at low frequencies. Blue dotted vertical lines indicate the positions of some previously reported periods (see Sect. 4.2 for details). {\it Middle:} Frequency spectra at high frequencies. The vertical dashed blue lines are placed at the Nyquist frequencies whereas the horizontal dashed red line indicates the 1\% significance level as determined from the formula of \citet{mah11}. {\it Right:} Close-up on the detected frequencies, with {\it SMEI} shown by the black solid lines, {\it BRITE} by the red dashed lines, and {\it TESS} by the blue dotted lines.}
\label{lcbz}
\end{figure*}

A temporal variance spectrum \citep[TVS, ][]{ful96} was built for these lines. It confirms the presence of highly significant variability over most of the line profile (as it is detected for velocities between --200\,km\,s$^{-1}$ and 300\,km\,s$^{-1}$, see Fig. \ref{pgram}). Only the most distant wings do not seem significantly affected. We then performed a period search using the same modified Fourier algorithm as before. Periodograms were built at each wavelength step and then averaged over the entire profile (Fig. \ref{pgram}). All examined lines are similar and revealed the presence of a periodicity at 12.7$\pm$0.1\,d$^{-1}$. A similar frequency is found when doing the period searches on line moments of order 0 ($EW$) and 2 (widths), rather than on the whole profile. This is likely a 1\,d$^{-1}$ (i.e. daily) alias of the photometric frequencies near 11.6 and/or 11.8\,d$^{-1}$. 

As this will be useful for interpreting the variability, we also derived the amplitude and phase of the best-fit sinusoid at each wavelength step using the modified Fourier algorithm as well as the usual least-squares fitting. This was done for both the spectroscopic (12.7\,d$^{-1}$) frequency and the strongest photometric (11.78\,d$^{-1}$) frequency. Both methods and both frequencies yield similar results: the amplitude reaches a maximum in the line core of 10--30\% of the line depth and the phase monotonically varies by 2.5--3 wave cycles across the profile (i.e. $\Delta\Psi_0=(5-6)\pi$ using the notation of \citealt{tel97}, see also Sect. 4.1). H$\alpha$ displays a similar behaviour, though the relative amplitude is only 3\%, because the emission component is much stronger than the photospheric absorption, and the phase difference across the profile only amounts to 4$\pi$. We repeated this fitting considering two frequencies, the main one and its second harmonic (i.e. twice the fundamental frequency). The results for the main frequency are unchanged; the second harmonic shows an amplitude of about one quarter to one half of the fundamental amplitude and a phase difference across the profile of $\Delta\Psi_1\sim6\pi$ (Fig. \ref{pgram}).

Finally, we analyzed in the same way the 2019 OHP monitoring of \pa. This time, the disk was much stronger hence the spectrum is dotted by numerous emission lines, notably from Fe\,{\sc ii}. Nevertheless, despite the higher level of disk activity at that time and the lower spectral resolution of the spectra, \pa\ still clearly showed fast migrating subfeatures in He\,{\sc i}\,$\lambda$4471, Si\,{\sc iii}\,$\lambda$4552, and H$\beta$ (Fig. \ref{ohp}). Their amplitude may at first seem smaller than in the 2006 data. However, we convolved the FEROS spectra with a Gaussian until we reach the spectral resolution of the OHP setting and the resulting amplitudes of the FEROS subfeatures are similar to those recorded in OHP data, demonstrating that the different appearance of greyscale plots is only due to the different resolutions. Analyzing the OHP data further, the TVS confirms the presence of significant variability over the whole profile (a velocity range between --250\,km\,s$^{-1}$ to 250\,km\,s$^{-1}$, Fig. \ref{ohp}). Period searches confirm the presence of a $\sim$12\,d$^{-1}$ signal, with similar amplitude/phase behaviour as in 2006 (Fig. \ref{ohp}). From these additional spectra, we conclude that the short-term variability of \pa\ does not appear to have profoundly changed despite large modifications in its disk.

\subsection{\bz}

\begin{figure}
  \begin{center}
\includegraphics[width=8.5cm,bb=50 5 700 740,clip]{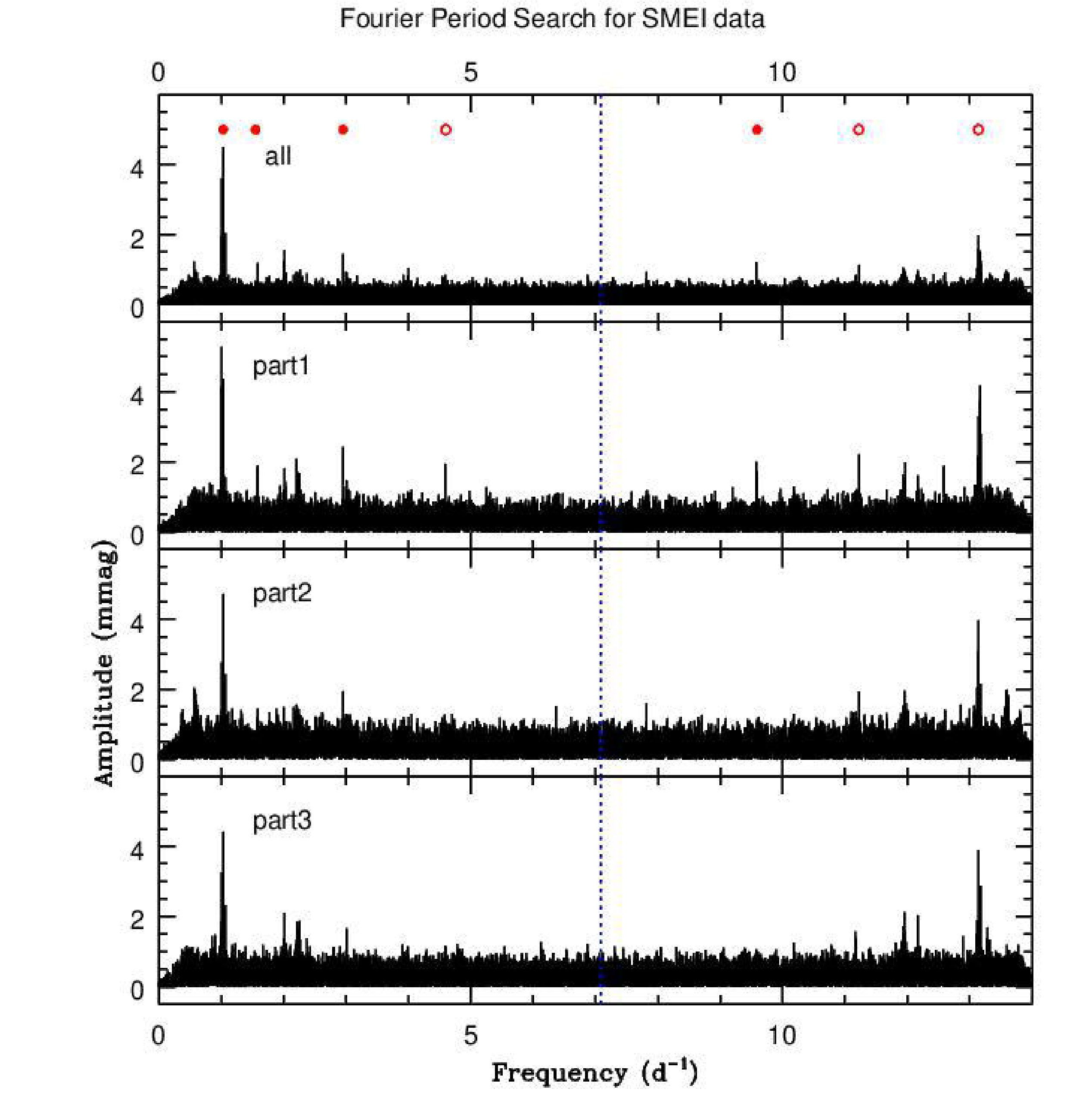}
  \end{center}
  \caption{Comparison of the frequency spectra derived by the modified Fourier method for the three parts of the {\it SMEI} light curve of \bz\ (with a detrending window of 0.3\,d). Red dots in top panel indicate the astrophysical frequencies mentioned in the text, red circles are used for their aliases; the vertical blue dotted line is placed at the Nyquist frequency. }
\label{3parts}
\end{figure}

\subsubsection{Photometric analysis}
Three photometric time series of \bz\ made from space are available (Fig. \ref{lcbz1}). They were analyzed with the same period search techniques as used for \pa\ (see Sect. 3.1.1). As a few long period candidates had been proposed for \bz\ (see Sect. 4.2 below), we first examined the light curves for their presence. For {\it SMEI} data, we used the raw light curve (i.e. cleaned but before any detrending) to this aim. For {\it TESS} and {\it BRITE} monitorings, we recall that they were limited to 24 and 78 days, respectively, hence cannot constrain periods longer than a few weeks. The results, for the modified Fourier technique\footnote{In addition to this method, we also used the conditional entropy and AOV methods on {\it BRITE} and {\it TESS} data (as for \pa, they are of little use for {\it SMEI} data). Their results confirm the frequencies found by the modified Fourier method, though the periodograms become complicated for {\it BRITE} because of the numerous subharmonics (unavoidable for these methods) as well as their aliases.}, are shown in the left panel of Fig. \ref{lcbz}. Because of the presence of some (limited) long-term variations, some signal is present at very low frequencies for the {\it SMEI} periodogram. It renders the detection of very long periods difficult, but no strong signal seems detected for the previously proposed long timescales of $\sim$130\,d \citep[][from optical observations]{smi06} or 226\,d \citep[][from X-ray observations]{smi12}. A peak is present in our {\it SMEI} periodogram at a frequency of 0.71462$\pm$0.00003\,d$^{-1}$ (or a period of 1.39935$\pm$0.00007\,d), which is close to a period previously proposed by \citet[see Sect. 4.2]{bar91}. However, it is very probably spurious. Indeed, the spectral window possesses a peak at a similar frequency. In addition, the peak disappears if detrending with long intervals is applied and, with its amplitude of 4.4\,mmag, it is associated with a significance level of 3\%, considering the formula of \citet{mah11}. Moreover, it is not obviously detected in either {\it TESS} or {\it BRITE} data although such a detection would not be prevented by their duration. Therefore we do not consider this signal as a real detection.

Turning to the highest frequencies and therefore using the cleaned and detrended light curves (middle panel of Fig. \ref{lcbz}), one frequency near 9.6\,d$^{-1}$ immediately stands out as it is detected by all three instruments. It appears with a frequency of 9.58800$\pm$0.00003\,d$^{-1}$ in the {\it SMEI} data cleaned by the first method, 9.58794$\pm$0.00003\,d$^{-1}$ in the {\it SMEI} data cleaned by the second method, 9.5840$\pm$0.0013\,d$^{-1}$ in the {\it BRITE} data, and 9.588$\pm$0.004\,d$^{-1}$ in the {\it TESS} data - the values being compatible within errors. Its amplitude varies between 1.2 to 1.9\,mmag, depending on the dataset under consideration. It appears to be highly significant as the worst significance level associated with it only reaches 0.005\%. This signal therefore is undoubtly stellar, and stable over long timescales. Note that the peak near 5\,d$^{-1}$, particularly detectable in the {\it BRITE} periodogram, corresponds to the alias of this 9.6\,d$^{-1}$ signal (considering the spacecraft's orbital period).

In parallel, we also detect a signal near 1.6\,d$^{-1}$: 1.57582 or 1.57586$\pm$0.00003\,d$^{-1}$ (for the two {\it SMEI} cleanings), 1.5731$\pm$0.0013\,d$^{-1}$ (for {\it BRITE}), or 1.576$\pm$0.004\,d$^{-1}$ (for {\it TESS}). Its amplitude reaches 1.2--1.5\,mmag. Again, the worst significance level associated with it reaches 0.005\%, making this signal highly significant. The same frequency simply seems detected in all datasets but the {\it TESS} periodogram also shows nearby companion peaks at 1.516\,d$^{-1}$ and 1.4\,d$^{-1}$, with the former one having the largest amplitude. This could thus point to a change in frequency between the older datasets and the most recent one: either this is a single signal whose frequency value changed over time or there are two independent frequencies with varying amplitudes (one being strong when the other one is weak). 

Additional frequencies are also present. In {\it SMEI} data, peaks occur near 1.03, 2.01, 2.95, 11.22, and 13.14\,d$^{-1}$. The latter two are aliases of the 2.95 and 1.03\,d$^{-1}$ frequencies, considering the spacecraft's orbital period. As mentioned previously, the other are (near) integer-frequency signals which are often measured in {\it SMEI} data \citep{gos11}, forbidding any real stellar contribution at these frequencies to be identified. However, this time, similar signals at 1.03--1.04 and 2.92--2.95\,d$^{-1}$ (with slightly displaced aliases because of the different orbital period of the spacecrafts) are also found in {\it BRITE} and {\it TESS} data, although these spacecrafts are not known to suffer from the same problem as {\it SMEI}. They are thus most probably real. No other frequency is detected - in particular, there is no signal beyond 14\,d$^{-1}$ (i.e. outside what is shown in Fig. \ref{lcbz}).

\begin{figure}
  \begin{center}
\includegraphics[width=8cm, bb=20 175 590 420, clip]{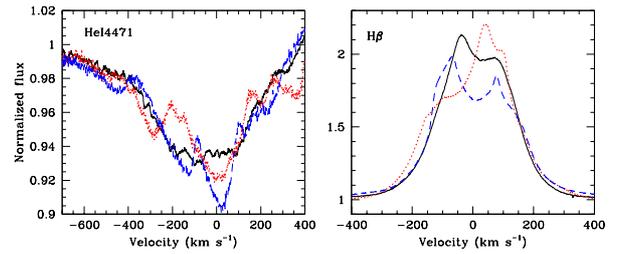}
  \end{center}
  \caption{He\,{\sc i}\,$\lambda$4471\,\AA\ and H$\beta$ line profile in UVES data of \bz\ in 2003 (black solid line), 2009 (red dotted line), and 2018 (blue dashed line). }
\label{specbz}
\end{figure}

\begin{figure*}
  \begin{center}
\includegraphics[width=5.cm, bb=0 110 390 730, clip, angle=90]{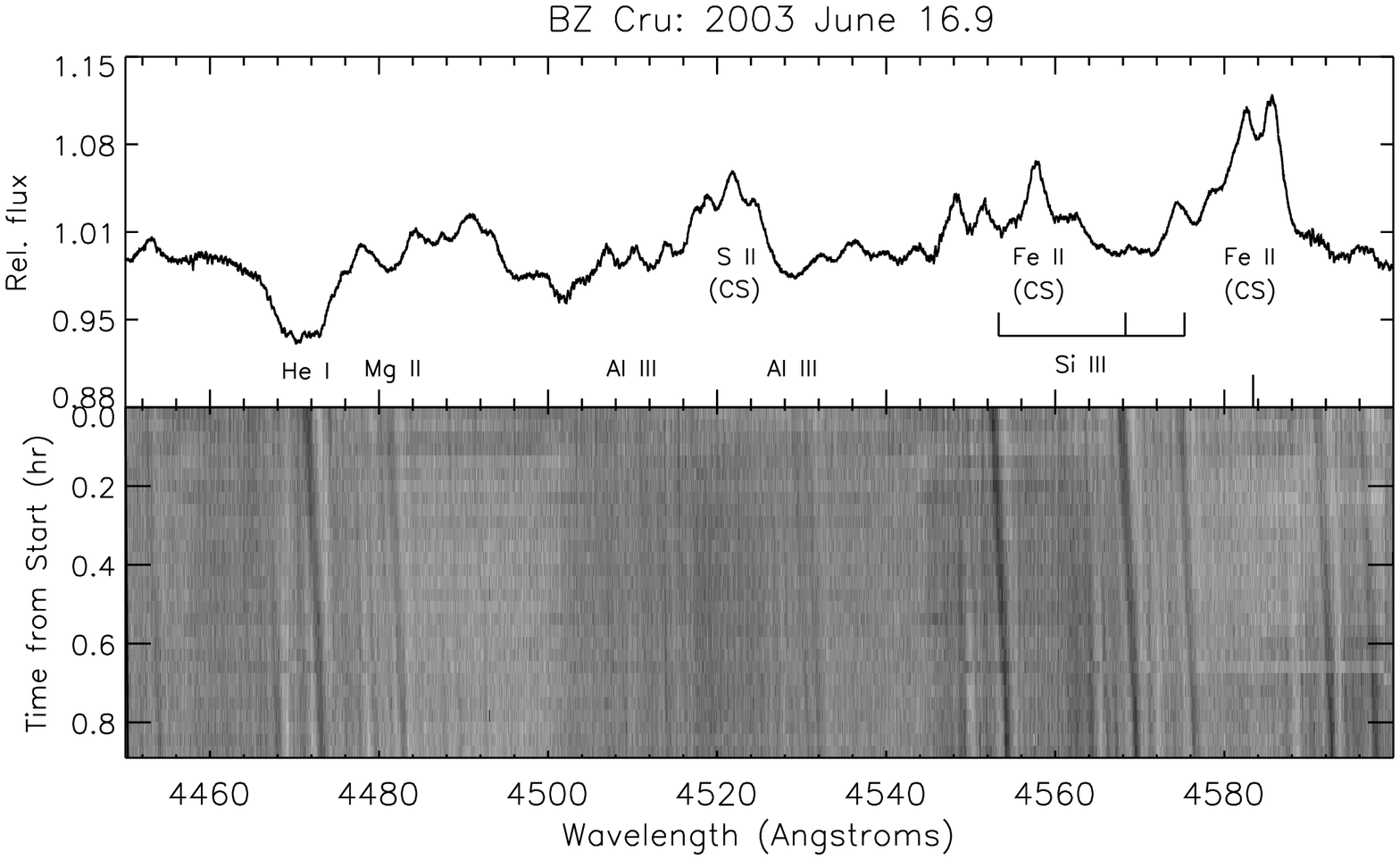}
\includegraphics[width=4.7cm]{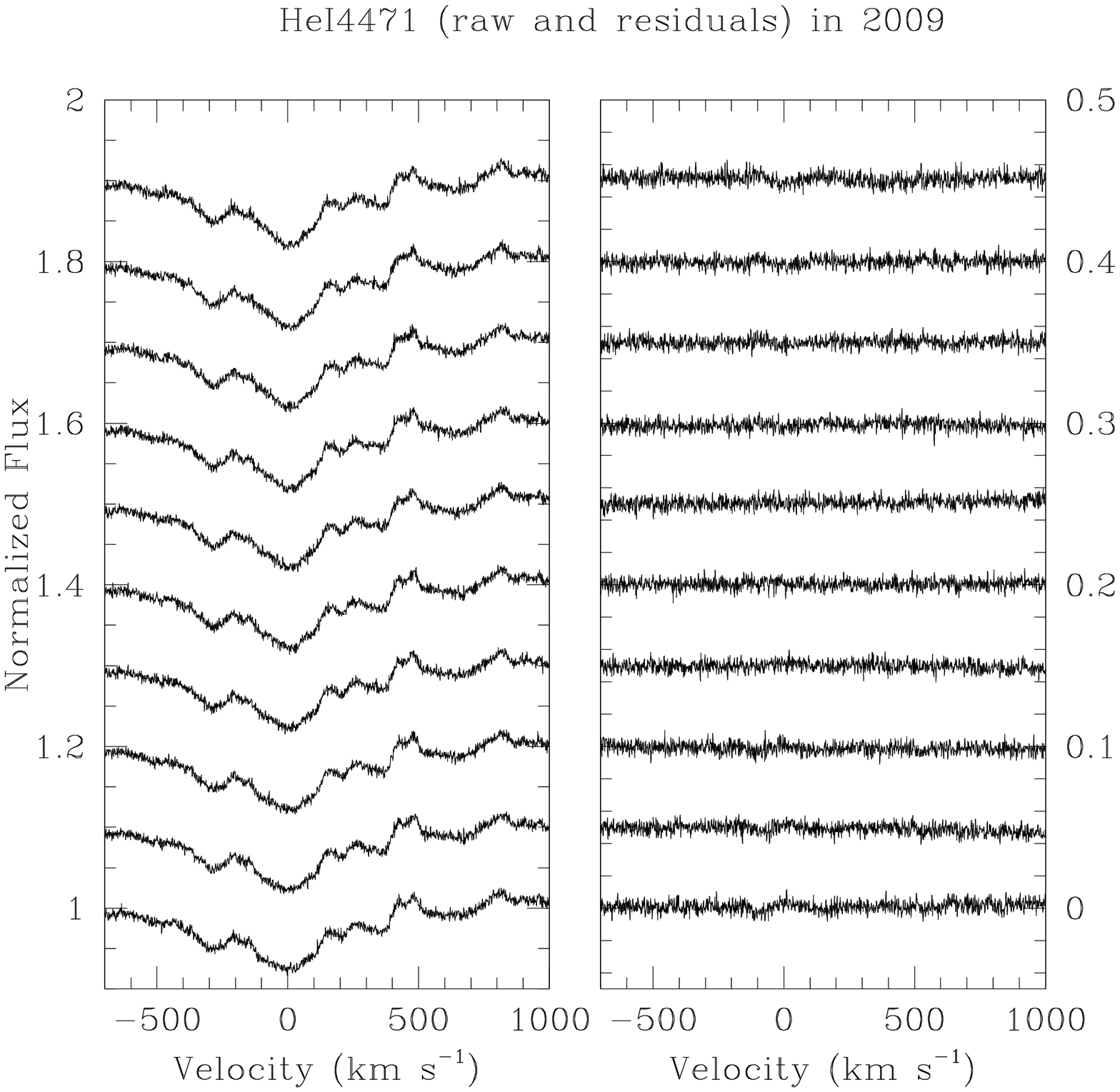}
\includegraphics[width=4.7cm]{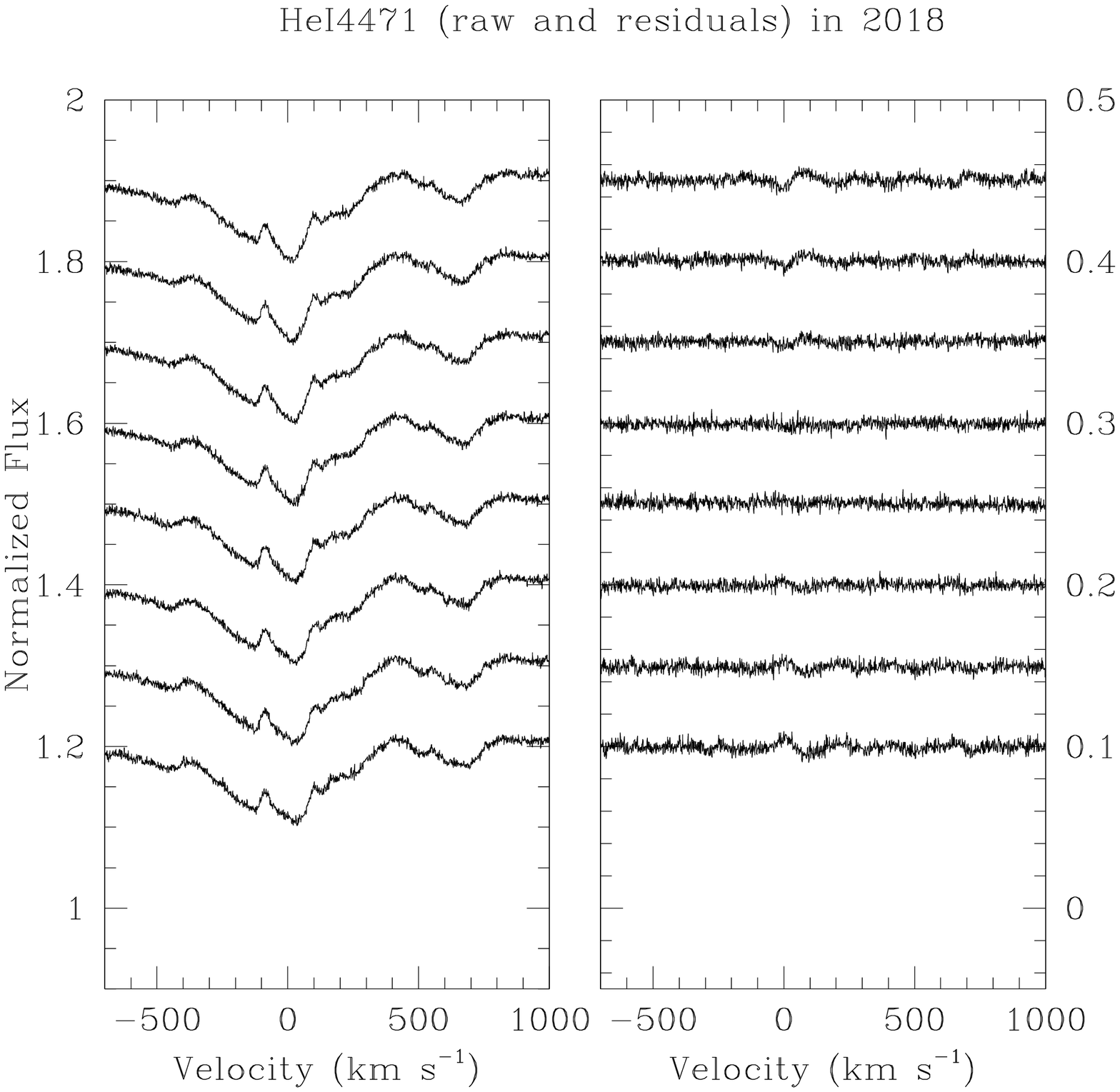}
  \end{center}
  \caption{Evolution of the line profiles of He\,{\sc i}\,$\lambda$4471\,\AA\ in UVES monitoring of \bz. {\it Left:} Top panel displays the mean UVES spectrum of \bz\ in 2003, with prominent lines identified (``CS'' indicate lines associated with circumstellar material). Bottom panel provides greyscale difference spectra for 2003 June 16 (last UVES night), with time running towards the bottom. {\it Middle and right:} In each subfigure (middle for 2009, right for 2018), the left panel yields the evolution of normalized line profiles while the right one provides their residuals (profiles minus mean). Individual profiles are arbitrarily shifted in the vertical direction to facilitate comparison, with time increasing towards the bottom. No significant variations are detected.}
\label{profbz}
\end{figure*}

\begin{figure*}
  \begin{center}
\includegraphics[width=5.8cm]{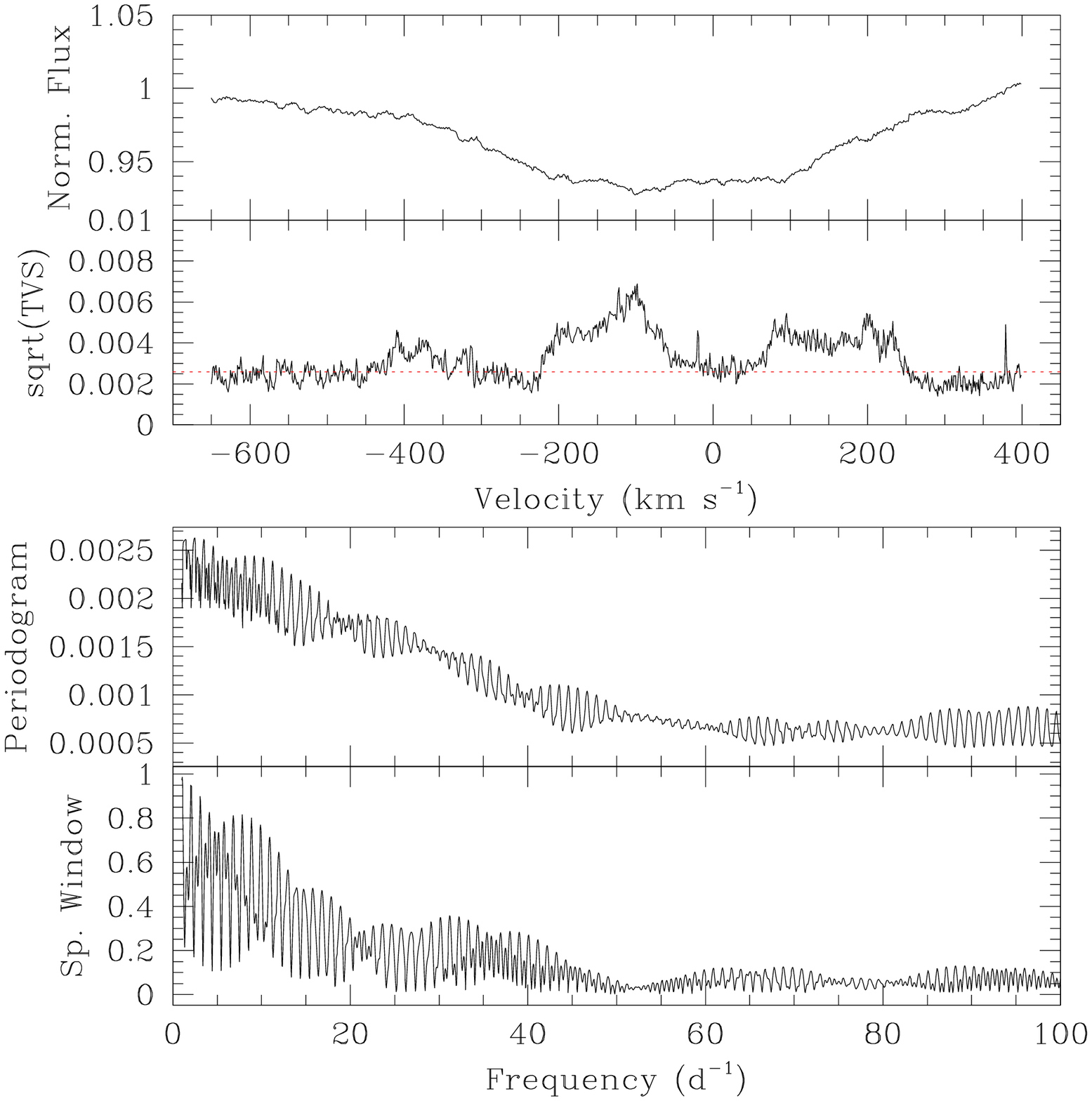}
\includegraphics[width=5.8cm]{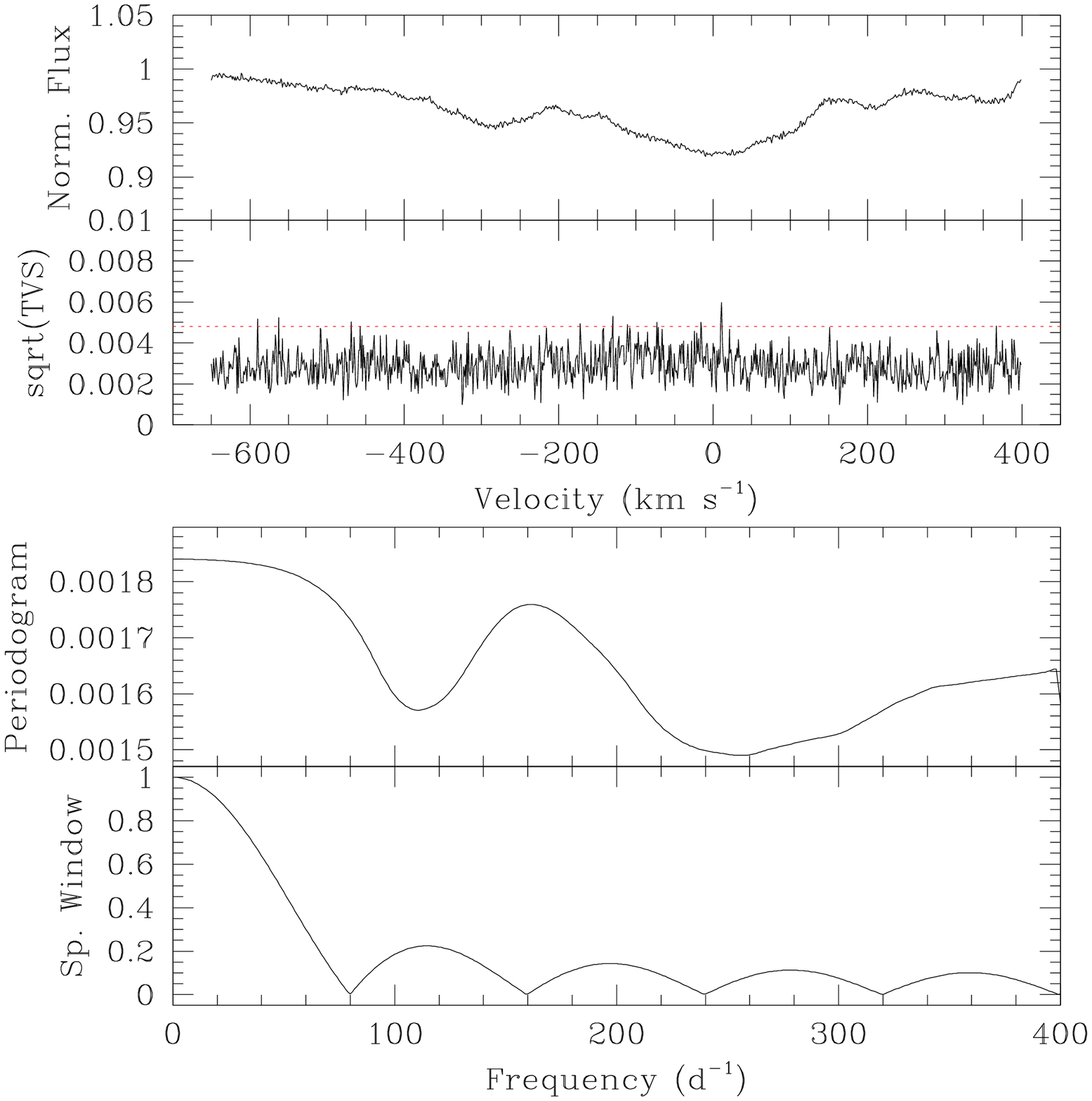}
\includegraphics[width=5.8cm]{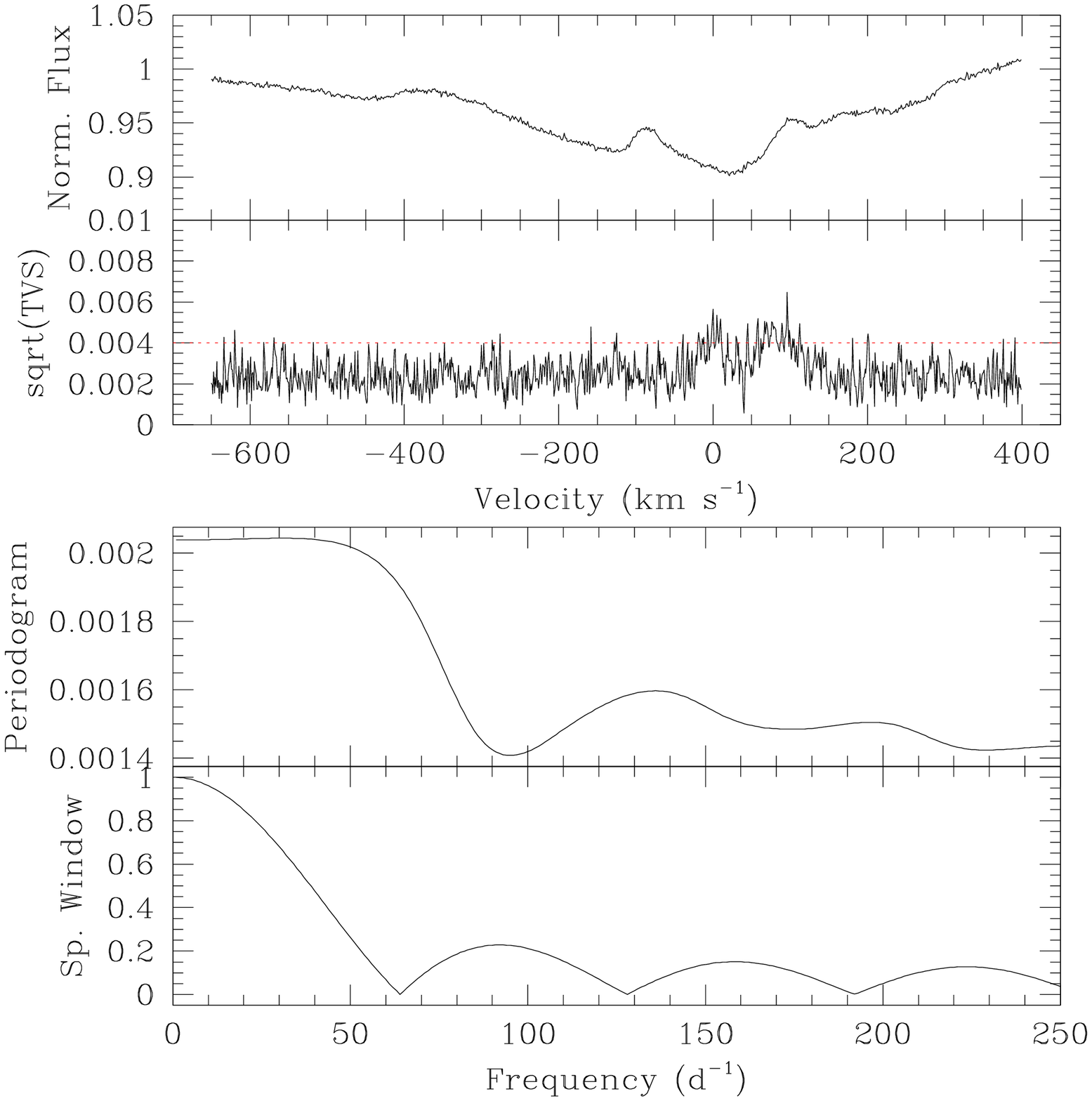}
  \end{center}
  \caption{Variability properties of He\,{\sc i}\,$\lambda$4471\,\AA\ in the UVES data of \bz. From top to bottom: mean line profile, TVS (with the 1\% significance level in red), periodogram in the same velocity range, and spectral window associated with the sampling for the 2003, 2009, and 2018 data (left, middle, and right, respectively).  }
\label{fourbz}
\end{figure*}

Finally, the undetrended {\it SMEI} light curve of \bz\ reveals the presence of long-term changes, though without multiple outbursts as seen in \pa\ (Fig. \ref{lcbz1}). The star first appears slightly brighter, while brightness dips are seen towards the end. We therefore decided to cut the {\it SMEI} light curve into three parts (Fig. \ref{lcbz1}), which we analyzed separately. The resulting periodograms, shown in Fig. \ref{3parts}, revealed that both the 1.6\,d$^{-1}$ and 9.6\,d$^{-1}$ signals (and their aliases) appear stronger in the first part of the light curve, becoming barely significant in the others. The 2.9\,d$^{-1}$ frequency is always clearly present, but also with a varying amplitude. In contrast, the 1.03\,d$^{-1}$ peak always appears strong, with more limited amplitude variations. Fitting a 4-sine model with frequencies fixed at 1.034, 1.576, 2.946, and 9.588\,d$^{-1}$ to the light curves confirms these evolutions. 

\subsubsection{Spectroscopy}
The 2003, 2009, and 2018 UVES spectra of \bz\ were all obtained as the star displayed strong Balmer emission lines ($EW$(H$\beta$)$\sim$--5\AA) and numerous small Fe\,{\sc ii} emissions, indicating a high level of disk activity. Because of this activity and the spectral ranges covered by the datasets, we decided to focus on two isolated lines which were repeatedly observed: He\,{\sc i}\,$\lambda$4471\,\AA\ and H$\beta$. In \pa, these two lines clearly showed the signature of variations, even when disk activity was high, hence they constitute good diagnostics. In addition, we examined Si\,{\sc iii}\,$\lambda$4552\,\AA: this line is lost amongst Fe\,{\sc ii} emissions but it still clearly showed short-term variations for \pa\ in 2019 when the disk lines were strong. On the other hand, we could not study in depth other potentially interesting He\,{\sc i} lines: He\,{\sc i}\,$\lambda$4921\,\AA\ suffers from imperfect order merging which causes some trouble in pinpointing the stellar variations; the line with the largest variations in \pa, He\,{\sc i}\,$\lambda$6678\,\AA, cannot be used as it was not covered by these UVES spectra. 

\begin{figure}
  \begin{center}
\includegraphics[width=8.cm]{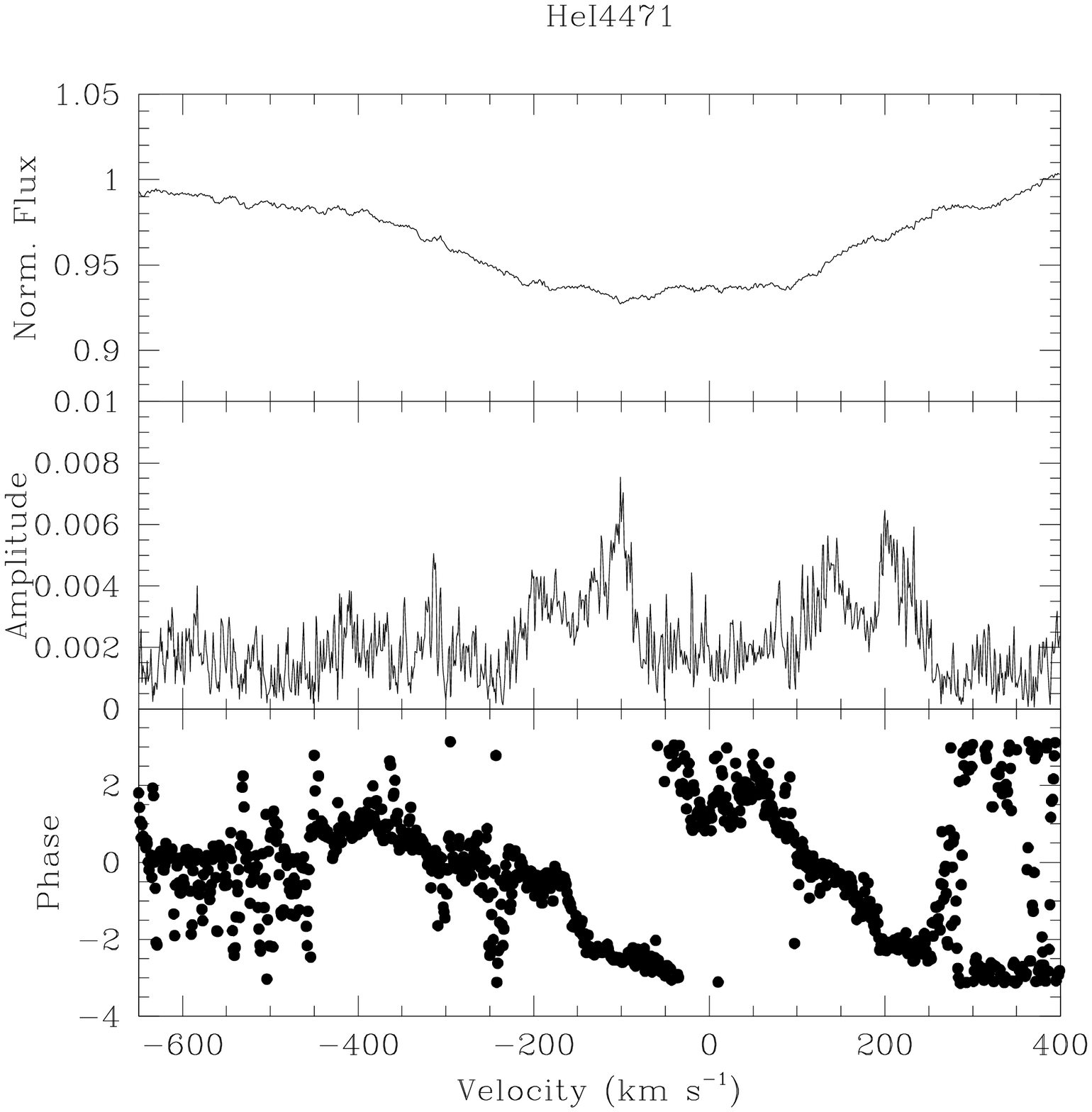}
  \end{center}
  \caption{From top to bottom: mean profile of the He\,{\sc i}\,$\lambda$4471\,\AA\ line, amplitude and phase of the best-fit sine wave (with $f=9.59$\,d$^{-1}$) for the 2003 UVES data of \bz. }
\label{fitsinbz}
\end{figure}

The first result from these datasets is that, while the disk lines always remain intense, long-term changes in line profiles exist (Fig. \ref{specbz}). They are most probably linked to variations in disk structure. In contrast, short-term variability does not appear to have a large amplitude, as shown in Figure \ref{profbz}. Considering the noise, migrating absorptions with full amplitudes of several per cent of the continuum level, as seen in \pa, would have been readily visible but they are not present. However, fast-moving, smaller-scale ($\sim$1\%) features existed in 2003. They appear with the same profile and position in the residual spectra of both He\,{\sc i}\,$\lambda$4471\,\AA\ and the silicon lines (Fig. \ref{profbz}), discarding noise or an instrumental defect as explanation for their presence. While the subfeatures move from blue-to-red with a similar rate as in \pa, the difference spectra in \bz\ remain however ambiguous as to the presence of periodicity. Another feature clearly appears on the greyscale difference spectra (Fig. \ref{profbz}): the absence of the migrating absorptions in several circumstellar emission lines such as e.g. the Fe\,{\sc ii}\,$\lambda$\,4584\,\AA. This tells us unambiguously that the migrating subfeatures are not linked to blobs in the disks of Be stars coming into the line-of-sight, as sometimes suggested \citep{har99}, implying that such disks are probably smooth in density on scales much smaller than the stellar radius.

Moments of lines were calculated as for \pa, but with the following velocity ranges: --650\,\kms\ to 400\,\kms\ for  He\,{\sc i}\,$\lambda$4471\,\AA, --325\,\kms\ to 325\,\kms\ for H$\beta$, and --3\,\kms\ to 13\,\kms\ for Na\,{\sc i}\,$\lambda$5895. Note that 53 spectra were used for 2003, 10 in 2009 and 8 in 2018 for He\,{\sc i}, 10 in 2009 and 2018 for Na\,{\sc i} lines, and 20 and 18 for H$\beta$ in 2009 and 2018, respectively. Table \ref{mombz} provides the average values with their observed dispersions. The larger dispersions for the full sample compared to individual campaigns confirm the presence of year-to-year variations, whereas the larger dispersion of stellar lines compared to interstellar ones shows that some limited short-term variability is also present. The situation is revealed in more detail by the calculation of temporal variance spectra (Fig. \ref{fourbz}): at most, very mild, barely significant variability is detected in 2009 and 2018 while, in 2003, clearly significant variability occurs over the velocity interval ranging between --200\,\kms\ and 200\,\kms\ for both surveyed lines (He\,{\sc i}\,$\lambda$4471\,\AA\ and Si\,{\sc iii}\,$\lambda$4552\,\AA). Finally, as in \pa\ (Sect. 3.1.2), we performed a period search using the modified Fourier algorithm, again averaging the periodograms calculated at each wavelength step over the entire profile (Fig. \ref{fourbz}). No clear periodicity is detected, but it must be remembered that the UVES data span about 2\,d, 0.02\,d, and 0.04\,d in 2003, 2009, and 2018, respectively. The two most recent spectral sets are thus only able to uncover periodic behaviour with very high frequencies (larger than tens of d$^{-1}$). Considering the detected photometric frequencies, the absence of significant variations in these data is thus not surprising, nor are the increased amplitudes at $f<100$\,d$^{-1}$ in their periodograms. The 2003 data are sparse but their sampling should in principle be sufficient to uncover at least the 9.6\,d$^{-1}$ signal. However, the periodogram remains ambiguous, hinting only at the presence of variations with ``low-frequency'' (i.e., $f<15$\,d$^{-1}$ in this case) timescales and not clearly leading to the detection of a periodic signal.

\begin{table}
\centering
\caption{Average line moments and their dispersion measured in three UVES monitoring runs of \bz. Rest wavelengths are as in Table \ref{mom} for He\,{\sc i} and H$\beta$, and 5895.924\AA\ for the interstellar Na\,{\sc i} line (marked with an asterisk, as in Table \ref{mom}). }
\label{mombz}
\setlength{\tabcolsep}{3.3pt}
\begin{tabular}{lcccc}
\hline\hline
Line & Year & $EW$(\AA) & $M_1$(km\,s$^{-1}$) & width(km\,s$^{-1}$) \\
\hline
He\,{\sc i}\,$\lambda$4471   & 2003 & 0.57$\pm$0.03     & --91$\pm$7   & 212$\pm$3       \\
He\,{\sc i}\,$\lambda$4471   & 2009 & 0.59$\pm$0.01     & --73$\pm$3   & 235$\pm$2       \\
He\,{\sc i}\,$\lambda$4471   & 2018 & 0.64$\pm$0.01     & --92$\pm$2   & 216$\pm$1       \\
He\,{\sc i}\,$\lambda$4471   & all  & 0.58$\pm$0.03     & --88$\pm$9   & 216$\pm$8       \\
H$\beta$                     & 2009 & --5.339$\pm$0.009 &6.9$\pm$0.4   & 122.3$\pm$0.2   \\
H$\beta$                     & 2018 & --4.728$\pm$0.007 &3.8$\pm$0.1   & 124.6$\pm$0.1   \\
H$\beta$                     & both & --5.1$\pm$0.3     &5.5$\pm$1.6   & 122$\pm$3       \\
Na\,{\sc i}\,$\lambda$5895$^*$& 2009 & 0.2023$\pm$0.0001 &5.32$\pm$0.01 & 3.380$\pm$0.005 \\
Na\,{\sc i}\,$\lambda$5895$^*$& 2018 & 0.2035$\pm$0.0001 &5.23$\pm$0.02 & 3.316$\pm$0.004 \\
Na\,{\sc i}\,$\lambda$5895$^*$& both & 0.2029$\pm$0.0006 &5.28$\pm$0.05 & 3.35$\pm$0.03   \\
\hline      
\end{tabular}
\end{table}

Nevertheless, since a clear high-frequency periodicity was detected in the photometry, we investigated whether the spectral variability detected in 2003 is compatible with it. As was done for \pa, we fitted a sine wave at each wavelength step considering a frequency of 9.59\,d$^{-1}$. Results are not as clear as for \pa: for the best case (He\,{\sc i}\,$\lambda$4471\,\AA, Fig. \ref{fitsinbz}), the amplitude reaches at most 10\% of the line depth and the phase varies by $\Delta\Psi_0\sim3\pi$ though its change appears somewhat incoherent/noisy across the profile. 

\begin{table}
\centering
\caption{Main photometric frequencies detected for our targets.}
\label{res}
\setlength{\tabcolsep}{3.3pt}
\begin{tabular}{lcc}
  \hline\hline
  Star & Ins. & $f$ (d$^{-1}$)\\
\hline
  \pa & SMEI  & 3.2979, 7.3332, 8.3194\\
      &       & several near 11.575\\
      &       & 11.7799,11.7807\\
  \bz & SMEI  & 1.0343, 1.5758, 2.9458, 9.5880\\
  \bz & BRITE & 1.039, 1.573, 2.918, 9.584\\
  \bz & TESS  & 1.032, 1.576, 2.952, 9.588\\
\hline
\end{tabular}
\end{table}

\section{Discussion}
In this section, we first place our results in the context of previous observations of each star, compare them with one another, and then discuss them in the general context of short-term variability in Be stars and in particular in $\gamma$\,Cas stars.

\subsection{\pa}
The variations of \pa\ have been studied for decades. Long-term spectroscopic changes of its H$\alpha$ emission, linked to the disappearance or built-up of its decretion disk, have been repeatedly reported \citep{mcl62,bjo02,zha13,naz19b}. Associated long-term photometric and polarimetric changes have also been observed \citep{fer75,nor77,gor85,bjo94}. No relationship between the orbital period of the system and the disk disappearance times could be found. In addition, \pa\ was found to show unusual variations in the photospheric components of the He\,{\sc ii}\,$\lambda$\,1640\,\AA\ and C\,{\sc iv}\,$\lambda\lambda$1548-50\,\AA\ lines \citep{smi06b}. Their variability timescale was long and the origin of this activity remains unknown.

The study of short-term variations yielded mixed results. \citet{hae75} performed a simultaneous spectroscopic, polarimetric, and photometric monitoring campaign over one week, but found no strong variations. Several authors reached the same conclusion \citep{sle78,fon83,gho88}, while others disagree, detecting night-to-night changes \citep{gra74}. While searching for non-radial pulsations in a set of Be stars, \citet{riv03} detected the presence of multiple sharp and fast transients on spectra of \pa, an unusual feature amongst their sample stars, but no periodicity could be established. Their small amplitudes may explain why they were not detected before. Finally, \citet{rin74} reported the detection of large velocity changes (from --100\,km\,s$^{-1}$ to 100\,km\,s$^{-1}$) arising with a period of 0.087\,d, or a frequency of 11.5\,d$^{-1}$. However, this finding was subsequently questioned \citep{fer75,hae75}. Indeed, monitorings rather found velocity changes of a much smaller amplitude (10--30\,km\,s$^{-1}$) and mostly linked to orbital motion \citep{bjo02,naz19}. More recently, in a proceedings, \citet{pet05} announced the detection of line profile variations occurring with a frequency of 12.8\,d$^{-1}$, consistent with our results. 

Our analysis of the {\it SMEI} photometry of \pa\ reveals long-term as well as short-term variations. Several short-term periodicities were detected, with frequencies 3.3, 7.3, 8.3, 11.6, and 11.8\,d$^{-1}$ (Table \ref{res}). Their amplitudes vary over time, with the 7.3 and 8.3\,d$^{-1}$ signal appearing stronger during one or both outbursts and the other signals appearing stronger mostly during ``quiescence'' episodes. The high frequency values would suggest p-modes of a high degree $l$, if the variability is interpreted as non-radial pulsations.

The FEROS spectroscopy of \pa\ taken in June 2006 clearly reveals the presence of a periodicity at 12.7$\pm$0.1\,d$^{-1}$, which most probably corresponds to the daily alias of the photometric frequency at 11.6--11.8\,d$^{-1}$. The same frequency was already present in 1999--2001 \citep{pet05}. At the time, the disk activity was at a low level, with a near-disapperance of emission lines \citep{bjo02}, while the disk activity was moderate at the time of the FEROS observations. Since then, the star has become much more active. The 2018 TIGRE optical-NIR spectra of \citet{naz19} sample three orbital cycles with one exposure taken every $\sim$10\,d. Therefore, they cannot be used for searching for high-frequency signals. However, if they were present, the affected spectral lines should be slightly different in each exposure. These data show little signs of strong line profile variability, except maybe for H$\alpha$ (though the presence of even stronger long-term changes prohibits a definitive conclusion) and for He\,{\sc i}\,$\lambda$6678\,\AA\ (which continued to show several sharp transient features). Indeed, for the last few years, the spectrum has been dominated by emission lines, hiding subtle profile changes at first sight. Nevertheless, a monitoring in 2019 at OHP, specifically dedicated to short-term changes, confirms the presence of low-level variability with a $\sim$12\,d$^{-1}$ frequency. As the same dominant frequency is present in several spectroscopic monitorings taken when the star was displaying very different disk activity levels, it suggests an overall stability of the signal, as well as an independence of the disk itself - i.e., it truly concerns the stellar photosphere.

If we consider photospheric pulsations as the source of variability, we may try to constrain their properties. Line-profile variability can indeed be used to constrain the mode caracteristics \citep[e.g.][]{tel97,zim06}. In particular, \citet{tel97} examined in detail the relation between the degree $l$ and azimuthal order $m$ of the pulsation and the phase difference across the line profile of the best-fit sine waves at the fundamental frequency and its second harmonic, respectively. In Sect. 3.1.2, we generally found $\Delta\Psi_0\sim(5-6)\pi$ and $\Delta\Psi_1\sim6\pi$. Using Eqs. (9) and (10) of \citet[see also their Eq. (4) and Fig. 1]{tel97} and inserting the quantities $\Delta\Psi_0$ and $\Delta\Psi_1$ determined in Sect. 3.1.2, we find $l\sim6\pm2$ and $|m|\sim2^{+2}_{-1}$ for this mode (which is prograde in view of the evolution of phase $\Psi$ across the profile). One caveat must however be mentioned. The \citet{tel97} method treated the rotation as a perturbation and used the velocity field from the first order perturbation theory. It is thus applicable only to the case when the pulsation frequency is much higher than the rotation frequency. Therefore, it should be applied with caution to fast rotators. For \pa, using the stellar parameters of \citet{bjo02}, the rotational period amounts to 1.2\,d. Since the pulsation frequency detected in spectroscopic and photometric data is $\sim$12\,d$^{-1}$, the ratio between pulsation and rotation frequencies is about 14, which is quite high. General asteroseismic modelling applicable to fast rotators begins to be performed, with the first steps towards similar diagnostics as in \citet{tel97} being recently made \citep[- see also pioneering models accounting for first-order effects of rotation in \citealt{tow97,mai03}]{ree17}. In this context, it may be remarked that the amplitude of the observed line profile changes peaks towards the line center whereas models often predict larger amplitudes in the wings \citep[e.g.][]{ree17} or at least a flatter amplitude distribution across the profile than observed in \pa. Certainly, specific asteroseismic modelling of \pa\ is now required, but this goes beyond the scope of this paper.

Our results indicate that the atmosphere of \pa\ is quite dynamic, particularly with respect to the tesseral mode tentatively associated with the $\sim$12\,d$^{-1}$ frequency. For linear adiabatic nonradial pulsations the warm and cool zonal variations across the star's surface tend to cancel - more and more efficiently with increasing degree $l$.  For B-star atmospheres the surviving contribution of these cancellations is only about 1 per cent for $l = 6$ \citep[and Dziembowski 1979, priv. comm.]{dzi77}. Hence our measured light amplitude of 0.001\,mag for \pa\ is actually indicative of a local fluctuation in flux of 0.001/0.01 or about 10\% at the phase of non-radial pulsation extremum. Then, assuming $T_{\rm eff}$=25\,kK and in the Rayleigh-Jeans approximation, this 10\% variation in flux would suggest a temperature modulation of roughly 2.5\,kK. Individual photometric contributions from many small surface zones are too small to be detectable. This is in contrast to the redistribution of spectral fluxes across line profiles (through Doppler imaging), which we found to be large enough to model attributes of these same pulsations and estimate their modal degree and order.

An example of this sensitivity is provided by the C\,{\sc ii}\,$\lambda\lambda$\,6578,\,6583\,\AA\ doublet. In the mean spectrum of \pa\ (see Fig. \ref{graysc}), these lines are not even detectable but, in difference spectra, their line profile variability appears very strong. Individual difference spectra reveal that this is due to noticeable velocity-separated absorption and emission components. Usually, these C\,{\sc ii} lines are unremarkable in their strength (i.e., they are weak in early B main-sequence stars) and they exhibit only modest variability. However, their intrinsic profile exhibits weak wings \citep{you81}, making this doublet suitable for studies of line profile variability.

A well known exception to the rule that this doublet does not change much in most variable early B stars is the pulsationally variable BW\,Vul \citep{you81}. The photosphere of this star undergoes strong periodic shocks, first emanating from the underlying envelope and shortly thereafter an additional shock from gas raining back from the previous cycle in nearly free fall. The $EW$ curves of each C\,{\sc ii} line remained at 250\,m\AA\ through most of the pulsation cycle, except during the passages of the two shocks through line forming regions, when they decreased to only 50\,m\AA. To investigate the cause of this dramatic variation, \citet{smi03} and \citet{smi05} examined the behaviour of the doublet and other lines in the spectrum by means of model atmosphere simulations. In searching for causes of these extreme $EW$ variations, \citet{smi03} were able to rule out changes in atmospheric microturbulence, density, and continuous opacities, but \citet{smi05} further showed that a change in the local temperature gradient through the atmosphere can cause the changes in $EW$. That is to say, the observed large $EW$ variations of this doublet in the BW\,Vul spectrum could only be simulated by artificially raising the temperature in the line forming regions relative to temperatures in much deeper layers by 2--3\,kK. In principle, shock heating of the upper atmospheric layers could also produce this result. The strength of this doublet in the \pa\ spectrum shows similar extraordinary variations, but this time probably from a high $l$ non-radial pulsation. These seem likewise to be somehow due to departures from a standard hydrostatic temperature distribution for sightlines from local surface elements to the observer. In assessing which surface regions of the star tend to contribute most to the doublet's variations, e.g., the disk center or areas near the limbs, it is  probably important that its variations are in phase with those (albeit weaker) variations of most other lines in the optical spectrum, as indeed in BW\,Vul \citep{you81}. This argues that the central disk regions provide most of the contributions and that the kinetic motions are largely vertical.

\subsection{\bz}
While \pa\ had been the subject of many investigations, this is not the case of \bz. This star only attracted more attention in recent years, after its hard, intense, and varying X-ray emission was discovered by \citet{tor01} and linked to $\gamma$\,Cas in \citet{smi06}. Many investigations have thus taken place in the high-energy domain \citep{tor01,tor12,lop07,smi12,tsu18}. A possible X-ray periodicity of 14\,ks had been proposed \citep{tor01} but subsequent X-ray datasets did not confirm it, demonstrating its spurious nature \citep{lop07,tor12}. \bz\ does however strongly vary at high energies, with short-term flare-like activity (like in $\gamma$\,Cas and \pa) and a possible 226\,d timescale \citep{smi12}.

In the optical, \citet{bar91} identified a period of 1.77\,d (or possibly 1.42\,d) and amplitude 20--50\,mmag from 34 measurements made with a photometer and spread over 10 nights. \citet{smi06} proposed a longer variability timescale of $\sim$130\,d (only one cycle, with amplitude 40\,mmag, was recorded in their data). The latter authors also reported fast-migrating, shallow absorption features in He\,{\sc i}\,$\lambda$6678\,\AA, but this spectral variability appeared only on some nights, which constrasts with the behaviour of \pa.

In the ultraviolet, \citet{cod84} reported the presence of narrow absorptions at very negative velocities. \citet{smi06} further showed that, while He\,{\sc ii}\,$\lambda$\,1640\,\AA\ remained stable, lines from Si\,{\sc iv}, C\,{\sc iv}, and N\,{\sc v} displayed large variations over 2--3\,d. Finally, \citet{ste13} studied the disk of \bz\ using interferometric techniques, finding it somewhat inhomogeneous but still following a Keplerian rotation. No trace of a companion was then detected. Contrary to \pa, \bz\ is thus not a known binary, though dedicated studies examining this question are scarce hence the question of binarity remains open.

Our photometry does not confirm the previously proposed (long) timescales but all photometric datasets agree on the presence of other, short-term periodicities (Table \ref{res}). A 9.6\,d$^{-1}$ frequency is notably detected with similar - but not identical - amplitudes in 2003--2011, 2016, and 2019. If attributed to pulsations, as was done for \pa, the putative phase changes $\Delta\Psi_0$ may suggest $l\sim3\pm2$ - note that for \bz, using the stellar parameters of \citet{ste13}, the rotation period is $\sim$0.7\,d hence the ratio between pulsation and rotation frequency would amount to $\sim$7. However, the two stars also present some differences in behaviour. \pa\ presented the vast majority of its photometric signals at high frequencies ($>$7\,d$^{-1}$), whereas \bz\ displays several low-frequency signals but only one high frequency. {\it If} the variability is interpreted in terms of non-radial pulsations, this mix of frequencies could indicate the presence of several g-modes and one p-mode. In addition, \pa\ showed large and persistent line profile variations, while \bz\ only displays small-scale, intermittent changes without a clear periodicity (the features move across the profile at a similar rate as those detected in \pa, though).

How different are \bz\ and \pa? In X-rays, \bz\ appears four times brighter but its spectrum is somewhat less hard \citep{naz18}. Regarding their stellar properties, \citet[and references therein]{naz18} reported effective temperatures of 28.5\,kK and 25.4\,kK, with bolometric luminosities $\log(L_{\rm BOL}/L_\odot)$ of 4.55 and 3.88, for \bz\ and \pa, respectively. This agrees well with the values $T_{\rm eff}$=25kK, $\log(g)$=3.9, and $\log(L_{\rm BOL}/L_\odot)$=4.1 found by \citet{bjo02} for \pa. Both would then lie, in the Hertzsprung-Russell diagram, in the $\beta$\,Cephei instability strip. On the other hand, \citet{arc18}\footnote{Data and analysis results also available on http://besos.ifa.uv.cl} recently revised the stellar parameters, making them closer to each other. Indeed, they suggested the same spectral type (B1V) for both stars and favored effective temperatures of 22\,kK and 24.5\,kK, stellar radii of 6.0 and 5.9\,$R_\odot$, and gravities $\log(g)$ of 3 and 3.4 for \bz\ and \pa, respectively. The latter temperatures seem however a little low for the UV spectra to show the observed He\,{\sc ii}\,$\lambda$\,1640\,\AA\ lines. In any case, the stars appear quite similar. Therefore, the origin of their variability differences remains unclear.

\subsection{Comparison with other stars}

Fast migrating subfeatures, such as those detected in the line profiles of our two targets, were first found by \citet{rob86} in the rapidly rotating, magnetic dK2e star AB\,Dor. They were attributed to the passage of corotating prominences or clouds that transit in front of the star. A similar phenomenon was discovered by \citet{yan88} in the optical lines of $\gamma$\,Cas. Such features were also found to be ubiquitous in rapid cadence, high signal-to-noise ultraviolet spectra of that same star by \citet{smi99} and \citet{smi19}. 

The models of non-radial pulsations by \citet{vog83} and \citet{kam88} for fast rotators and \citet{tel97} for slow rotators showed that they can be a source of moving bumps in addition to absorptions from corotating clouds. In the profiles of a rapidly rotating star, these features would accelerate across the profiles at a rate similar to those associated with a corotating cloud. Therefore, the term ``migrating subfeatures'' might be expanded to include this scenario.

These two causes may nevertheless be distinguished from their characteristics. On one hand, corotating clouds exhibit the following attributes:
\begin{enumerate}
\item[a)]{the subfeatures are entirely absorptions relative to a mean profile created from a long spectral time series;}
\item[b)]{if more than one migrating subfeature pattern appear, they appear at irregular time intervals (as the underlying putative magnetic spots are not regularly separated in stellar longitude);}
\item[c)]{migrating subfeatures are short lived (up to a few rotations in AB\,Dor stars);}
\end{enumerate}
On the other hand, the tesseral non-radial pulsation modes are manifested by the following characteristics:
\begin{enumerate}
\item[i)]{they have a wave-packet appearance in difference spectra, i.e. separated quasi-emission lobes - in a limited time series not longer than the passage of a wave packet across a reference wavelength, the reflex of absorptions from an adjacent absorption lobe can cause a spurious emission-like feature (this sets a requirement for a lengthy time series);}
\item[ii)]{the wave packets are regularly spaced in time, unless influenced by beating of another mode closely spaced in frequency. The regular spacing of the spectral features can be matched with contemporaneous photometry, if it is available.}
\end{enumerate}
The $\gamma$\,Cas line profiles clearly pass the a--c criteria for corotating clouds \citep{smi19}. The spectra of \pa\ clearly pass at least the i--ii criteria but none of the a--c criteria. Because of the sparseness of the current spectral data and the shallowness of the features, the case of \bz\ is more ambiguous.  

We may also compare the derived photometric properties of \pa\ and \bz\ with previously published high-cadence photometry studies. Amongst $\gamma$\,Cas analogs, only the variability of $\gamma$\,Cas itself has been studied in detail before the present work. The prototype star displays a long-term photometric variability with timescale of 50--90\,d as well as longer-term, irregular variations \citep{hor94,smi12}. A signal with a 1.2\,d period was detected from 15\,yrs of {\it APT} data and 7\,yrs of {\it SMEI} data \citep{hen12,baa17}, whereas {\it BRITE} observations further indicated signals near 5.1, 2.5, 1.3, 0.97, and 0.82\,d$^{-1}$ (\citealt{baa17}, Borre et al. in prep). Contrary to \pa\ and \bz, no high ($>>5$\,d$^{-1}$) frequency signal was reported up to now for $\gamma$\,Cas, its {\it BRITE} periodogram actually being described as ``normal'' (amongst Be stars) by \citet{baa17}. As mentioned above, several authors reported the presence in the spectrum of $\gamma$\,Cas of absorption features migrating across the line profile at a rate of $\sim$2000\,km\,s$^{-1}$\,d$^{-1}$ \citep{nin83,yan88,hor94,smi95}, though no formal periodicity was derived for them hence their relationship (if any) to the photometric signals remains unknown.

The three $\gamma$\,Cas analogs studied up to now therefore present a range of behaviours. How do they compare to other Be stars? Be stars are known to be variable light sources, on both short and long timescales. The advent of space facilities dedicated to photometry such as {\it SMEI, MOST, CORoT, BRITE, Kepler,} or {\it TESS} led to a renewed view of the short-term photometric variability in Be stars. Generally, the periodograms display several (regularly-spaced) groups of (regularly-spaced) frequencies located below 5\,d$^{-1}$ \citep{cam08,gos11,bal11,riv16,baa18b,sem18,bal19}. Trying to get an overall view, \citet{bal19} analyzed a large sample of 57 Be stars observed with {\it TESS}. Three-quarters of their sample display low frequency signals, often gathered into groups in harmonic relation with each other. Such low-frequency signals are usually interpreted as g-modes typical of slowly pulsating B-stars (SPBs) though gravito-inertial modes have also been proposed in one case \citep{nei12} and a mix of p and g modes may occur for the earliest types \citep[$\zeta$\,Oph, ][]{cam08}. \citet{bal19} put forward an alternative scenario in which they hypothesize photospheric gas clouds and starspots leading to non-coherent variations recurring on the rotational period.

In the sample of \citet{bal19}, 19\% of Be stars display high-frequency ($f>5$\,d$^{-1}$) signals but only 5\% show high frequencies without low-frequency groups. With its few isolated high frequencies, \pa\ thus clearly appears different from the majority of the studied Be stars. It mostly resembles $\zeta$\,Oph \citep{wal05}, V739\,Mon (=HD\,49330, \citealt{hua09}), and HD\,58978 \citep{bal19}. These stars share a high effective temperature which places them well in the $\beta$\,Cep locus. Additional examples of high-frequency variations are found in the slightly cooler stars 27\,(EW)\,CMa \citep{bal91} and HD\,209014 \citep{bal19}. This similarity is confirmed by the spectroscopic analysis of these stars which yields $l=3-7$ \citep[and this work]{wal05,hua09}, in stark contrast with the $l=|m|=2$ (retrograde) modes found for the line profile variations of a majority of Be stars \citep{riv03}. 

In Be stars, the frequency content and the amplitude of the detected photometric signals were found to change with time \citep[e.g.][]{gos11,nei12,riv16,baa18}. Finding such changes in \pa\ or \bz\ is thus not surprising but details may vary from star to star. For example, \citet{hua09} showed that the frequency content of V739\,Mon changes during a small outburst (with $\Delta(mag)=0.03$). The p-modes became fainter whereas groups of g-modes appeared at low (i.e. $1<f<5$\,d$^{-1}$) frequencies. In \pa, outbursts recorded by {\it SMEI} are about ten times larger than that of V739\,Mon, and their properties are slightly different. While the dominant signals ($\sim$12\,d$^{-1}$) also became fainter during outbursts, the signals that then appear (7.3 and/or 8.3\,d$^{-1}$) are located at high ($f>5$\,d$^{-1}$) frequencies. The scenario of g-modes triggering the outbursts thus do not seem applicable to \pa. Variation of mode amplitude during outbursts in Be stars could also come from veiling of the equatorial regions by the flaring disk, especially for modes with $l=|m|$ as their amplitude is maximum at equator \citep[e.g.][]{mai03,bal19}. However, in \pa, the $\sim$12\,d$^{-1}$ signal was detected in spectra taken in 2006 and 2019, when the disk was in very different states, suggesting little influence of disk growth. Furthermore, a general extinction of all high-frequency signals (p-modes) during outbursts, as reported for V739\,Mon \citep{bal19}, is not detected in \pa. And for \bz, the sole high-frequency signal even appears stronger when the star is brighter (i.e. in the first part of the {\it SMEI} light curve). It thus also appears difficult to explain the amplitude changes of high-frequency signals observed in \pa\ and \bz\ as a consequence of obscuration by circumstellar matter ejected during the outburst.  Dedicated modelling is now needed to understand the different behaviours of all these stars.

\section{Summary and conclusions}
In this paper, the short-term variabilities of two $\gamma$\,Cas stars, \pa\ and \bz, have been examined. To this aim, we analyzed both ground-based spectroscopic campaigns and space-based photometric monitorings.

For \pa, the {\it SMEI} photometry reveals the presence of several high frequencies: 3.3, 7.3, 8.3, 11.6, and 11.8\,d$^{-1}$. There is no obvious frequency group nor any equidistant frequencies, as seen in other Be stars. The amplitudes of the detected signals vary with time, some being stronger during outbursts while others are dominant during quiescence episodes. In parallel, two intensive spectroscopic campaigns indicate the presence of strong line profile variations with an amplitude of several per cent of the continuum level and a $\sim$12\,d$^{-1}$ frequency, in agreement with the main photometric periodicities. The spectroscopic datasets were obtained when the disk of \pa\ was in quite different states, with a much more extensive disk during the last observing run. Therefore, the $\sim$12\,d$^{-1}$ signal seems not only long-lived but also independent of the disk hence truly stellar. The spectroscopic changes are consistent with non-radial pulsations, i.e. a tesseral mode having a high degree $l$, a quite unusual feature amongst Be stars (which are mostly SPBs) but which remains possible in view of the rather high temperature of \pa, which places it in the $\beta$\,Cep instability strip.

For \bz, {\it SMEI, BRITE}, and {\it TESS} photometric time series indicate the presence of several periodicities at 1.03, 1.52--1.58, 2.95, and 9.6\,d$^{-1}$. The amplitudes of these signals vary with time, especially those at 1.6 and 9.6\,d$^{-1}$. The spectra of \bz\ were all recorded when the disk was well developed - but not in perfectly identical states as long-term changes of the line profiles are ubiquitous. Spectroscopically, short-term variability appears to be of a much smaller amplitude than in \pa. In the 2003 data, some fast-moving, absorbing features are indeed detected but with amplitudes $<1$\% of continuum level. Furthermore, no clear periodicity can be associated with them, which contrasts with the results on \pa. 

In recent years, several Be stars were the targets of intense photometric monitorings. Usually, they show periodicities only at low frequencies ($f<$5\,d$^{-1}$) and appearing in regularly-spaced groups. This is not the case for \bz\ nor \pa, for which detected frequencies appear isolated and often at high frequencies. Whether this difference in frequency content is due to the high effective temperature or another physical property of these stars remains to be ascertained. Finally, the amplitudes of the periodic signals were found to change in some Be stars and in this respect, both \pa\ and \bz\ are no exceptions.

The exploration of the short-term behaviour of $\gamma$\,Cas analogs has only begun. Indeed, further advances in the field will require more Be stars, and in particular more $\gamma$\,Cas stars, to be observed in high-cadence, high-quality campaigns combining simultaneous spectroscopic and photometric observations. The derived observational properties should then be used for detailed theoretical modelling in order to assess whether $\gamma$\,Cas analogs display some specific characteristics making them differ from the other Be stars.

\section*{Acknowledgements}
Y.N. and G.R. acknowledge support from the Fonds National de la Recherche Scientifique (Belgium), the Communaut\'e Fran\c caise de Belgique (for financial support of the OHP campaign), the European Space Agency (ESA) and the Belgian Federal Science Policy Office (BELSPO) in the framework of the PRODEX Programme (contract XMaS). AP acknowledges support from NCN grant no. 2016/21/B/ST9/01126. We thank D.R. Reese for interesting discussions. ADS and CDS were used for preparing this document. This paper is notably based on data collected by the TESS mission, whose funding is provided by the NASA Explorer Program, and by the BRITE Constellation satellite mission, designed, built, launched, operated and supported by the Austrian Research Promotion Agency (FFG), the University of Vienna, the Technical University of Graz, the University of Innsbruck, the Canadian Space Agency (CSA), the University of Toronto Institute for Aerospace Studies (UTIAS), the Foundation for Polish Science \& Technology (FNiTP MNiSW), and National Science Centre (NCN).

\bsp	
\label{lastpage}
\end{document}